\documentclass{article}
\usepackage[a4paper, total={6in, 8in}]{geometry}
\usepackage[numbers,sort&compress]{natbib}
\usepackage{url}
\usepackage{physics}
\usepackage{graphicx}
\usepackage{caption}
\usepackage{subcaption}
\usepackage{mathrsfs}
\usepackage{amsmath,amssymb}
\DeclareMathOperator*{\argmax}{arg\,max}
\usepackage[utf8]{inputenc}
\usepackage[english]{babel}

\title{The Spherical Grasshopper Problem}

\author{Boris van Breugel\\
\textit{Department of Applied Mathematics and Theoretical Physics},\\\textit{University of Cambridge}}

\date{April 26, 2019}
\begin{document}
\maketitle
\abstract{
The aim of this essay is to better understand the Grasshopper Problem on the surface of the unit sphere. The problem is motivated by analysing Bell inequalities, but can be formulated as a geometric puzzle as follows. Given a white sphere and a bucket of black paint, one is asked to paint half of the sphere, such that antipodal pairs of points are oppositely coloured. A grasshopper lands on the sphere, and jumps a fixed distance in a random direction. How should the sphere be coloured such that the probability of the grasshopper landing on the same colour is maximized? Goulko and Kent have explored this problem on the plane without an antipodality constraint \cite{goulkokent}. This essay gives clear indication that the spherical problem with the antipodality constraint yields colourings with  similar shapes as the planar problem does.

This research has discretised the problem and used a simulated annealing algorithm to search for the optimal solution. Results are consistent with the planar results of \cite{goulkokent}. For $0.10\pi\leq\theta\leq0.44\pi$ cogwheel solutions are found to be optimal, with odd integer of cogs $n_o$ such that $n_o$ is close to $\frac{2\pi}{\theta}$. For $0.45\pi\leq\theta\leq 0.55\pi$ \textit{critical} solutions are found, in which domains of identical colour decrease in size towards $0.5\pi$ (moving from either side). For $\theta\geq 0.55$ colourings are found consisting of stripes with cogs. Towards $\theta=\pi$ colourings are generated that display just stripes that scale in width with $\pi-\theta$.
}





\section{Introduction}
This research focuses on the Grasshopper Problem  \cite{goulkokent} on the surface of the unit sphere. The problem is motivated by Bell inequalities. This section will briefly outline the quantum mechanical context, after which the Grasshopper Problem is introduced.

\subsection{Bell inequalities}
In June 1926 Max Born published a probabilistic interpretation of quantum mechanics in his paper \textit{Quantum Mechanics of Collision Phenomena}. Multiple leading scientists refuted this idea. Most notable is Einstein's famous reply to Born:
\begin{quote}
    Quantum mechanics is very worthy of regard. But an inner voice tells me that this is not yet the right track. The theory yields much, but it hardly brings us closer to the Old One's secrets. I, in any case, am convinced that \textit{He} does not play dice. \cite{einstein1971born}
\end{quote}

This criticism led to hidden-variable theories: the idea that there are underlying physical principles that deterministically govern the seemingly probabilistic quantum behaviour. Einstein, Podolsky and Rosen (EPR) argued for the incompleteness of quantum mechanics, and proposed a causal local hidden-variable theory (LHVT)\cite{LHVT}.

In 1964 John Bell published his famous article \textit{On the Einstein Podolsky Rosen paradox}\cite{bell}. In this work Bell inequalities are first introduced. Quantum theory predicts that space-like separated experiments performed on entangled particles can result in outcomes whose correlations would violate these Bell inequalities, whereas the inequalities would have been satisfied if the experiments could be described by an LHVT. Simply put, these inequalities provide an experimental method of determining whether the fundamentals of our world are dictated by either some classical mechanism or the probabilistic theory of quantum mechanics.

Bell inequalities have been used intensively in experimental research to provide proof that quantum mechanics is necessary for understanding the smallest scales of physics. Bell inequalities rely on space-like separated experiments that are difficult to perform in real-life. As a result, a number of notorious loopholes exist: imperfect characteristics of real-life experiments that could invalidate results. Most notable are the locality loophole, detection efficiency loophole and collapse locality loophole. Even taking into account the error bounds resulting from these imperfections, experiments \cite{prove1,prove2,prove3,prove4,prove5} have tested the quantum prediction of non-local causality and the resulting violation of Bell inequalities is in line with quantum mechanics. Thus, the existence of LHVTs can be refuted with high certainty.

Most experiments are based on simple Bell inequalities like the CHSH  (Clauser-Horne-Shimony-Holt) inequality \cite{CHSH} with an EPR-Bohm experiment set-up \cite{LHVT,bohm1951quantum}. However it would be valuable for extended research into more general Bell inequalities to be performed, to allow for a greater insight into the world of quantum mechanics, in particular quantum nonlocality. A possible application would be in the field of quantum cryptography. Quantum cryptographic protocols may provide a way of safe communication that can guarantee to give users notice if malevolent eavesdroppers or device manufacturers are after their sensitive data. One of the current challenges of quantum cryptographic protocols is efficiency; we want malicious parties to be detected within a reasonable number of tests. As these tests often rely on the use of Bell inequalities \cite{garciakent}, it is of great relevance to further explore the full class of Bell inequalities.

Although this essay will leave the context of quantum mechanics shortly, at this point it is worth sketching the idea of the quantum mechanical experiment that underlies the essay. The Bell inequality explored in this essay is based on the EPR-Bohm experiment. The experiment consists of two observers, Alice and Bob, who possess an entangled singlet state. Alice and Bob independently choose measurement axes $\mathbf{a}$ and $\mathbf{b}$ respectively with the constraint that the angle between the axes is $\theta$. They perform space-like separated measurements, and their outcomes are in $\{+1,-1\}$. The correlation $C(\mathbf{a},\mathbf{b})\in \left[-1,+1\right]$ is defined as the expected product of their outcomes.

Quantum mechanics predicts a correlation $C^Q(\theta)=-\cos(\theta)$.\footnote{Where with a slight abuse of notation we define $C(\theta)=C(\mathbf{a},\mathbf{b})$, with $\theta$ the angle between $\mathbf{a}$ and $\mathbf{b}$.} A valid Bell inequality would be either finding a lower bound $C^L(\theta)$ or upper bound $C^U(\theta)$ on the LHV correlation $C^{LHVT}(\theta)$ such that either $C^{LHVT}(\theta)\leq C^U(\theta) < C^Q(\theta)$ or $C^Q(\theta)< C^L(\theta)\leq C^{LHVT}(\theta)$. The key difference between this inequality and the standard CHSH inequality is that that observers Alice and Bob are free to choose their axes of measurements, with the constraint that the axes are separated by an angle $\theta$. For the standard CHSH experiment, Alice and Bob are given one axis, and the only freedom they have is choosing the direction in $\{+1,-1\}$ along which to measure. The new inequality is hence a generalisation of the CHSH inequality.  

\subsection{The Grasshopper Problem}
\citet{garciakent} translates finding tight bounds for the LHVT correlation to a geometric problem called the Grashopper Problem.  Informally this problem can be stated as follows. We are given a white sphere, a bucket of black paint and the task to paint one of every pair of antipodal points.\footnote{In the formulation of \cite{goulkokent} a lawn is used instead of black paint, but this metaphor does not translate well to the sphere with antipodality constraint.} A grasshopper lands on the sphere, and subsequently jumps a fixed angle $\theta$ in a random direction. The problem is, what colouring of the sphere maximises the probability that the grasshopper lands on the same colour after hopping, and what is this maximum probability as a function of $\theta$? 

In relation to the quantum context, the expected success probability $P(\theta)$ of the grasshopper is related to the bounds on the LHV correlation $C(\theta)$ for a specific case.\footnote{Specifically the case that there is perfect anticorrelation between the outcomes of Alice and Bob if they use the same measurements, $\theta=0$ (see \cite{garciakent}).} The relation between the grasshopper's success probability and the LHV correlation is $C(\theta)=1-2P(\theta)$. Consequently, finding upper and lower bounds on $P(\theta)$ in the Grasshopper Problem provides bounds on the LHV correlation. The rest of this essay will leave the context of quantum mechanics and focus on the Grasshopper Problem. Goulko and Kent \cite{goulkokent} have explored the planar problem. Inter alia, they have shown that a disc is not the optimal solution and have found numerical approximations for colourings for a variety of jumping distances. As the planar problem does not account for the antipodality condition, it is unclear whether the found solutions can be easily transferred to the spherical problem.  

This essay investigates the spherical Grasshopper Problem numerically. In Section 2, the problem will be formulated formally. In Section 3 the numerical set-up is discussed, including discretisation of the problem and approximation of the global maximum. In Section 4 the results are presented. In Section 5 these findings are discussed further, recommendations are made for improving the method, and a range of related problems is displayed.

\section{Problem statement}

The $\mathbb{S}^2$ Grasshopper Problem can be formally stated as follows. 

\subsection{Formal problem statement}
Consider a density $\mu$ on the surface $\mathbb{S}^2$ of the three-dimensional sphere, satisfying $0\leq\mu(\mathbf r)\leq 1$ and the antipodality condition $\mu(\mathbf r)+\mu(-\mathbf r)=1$ for all $\mathbf r$. Consequently: $$\int_{\mathbb{S}^2}d^2\mathbf{r}\mu(\mathbf r)=2\pi.$$
The functional $P_\mu(d)$ is defined by
\begin{align}
    P_\mu(\theta)=&\int_{\mathbb{S}^2}d^2r_1\int_{\mathbb{\mathbb{S}}^2}d^2r_2\mu(\mathbf r_1)\mu(\mathbf r_2)\delta(|\mathbf{r}_1-\mathbf{r}_2|-\theta),
\end{align}
where $|\mathbf{r}_1-\mathbf{r}_2|$ is defined as the central angle between points $\mathbf{r}_1$ and $\mathbf{r}_2$. The Grasshopper Problem is as follows:
$$\text{Given }\theta\text{, find}\argmax_{\mu} P_\mu(\theta)\text{ and}\max_{\mu} P_\mu(\theta).$$
The density $\mu$ denotes the colouring of the sphere. This essay explores the case where $\mu$ assumes as value either $0$ or $1$.\footnote{For the discrete case this is not a constraint on the success probability, as any non-binary colouring can be transformed into a binary colouring with a equal or higher success probability. A proof is given in Appendix \ref{sec:appendix}.}  

\subsection{Preliminary analytic results}\label{sec:premresults}
This problem is very difficult to solve and the probability can only be calculated analytically for very simple colourings. The easiest colouring is a hemisphere colouring, for which the success probability $P_{hem}(\theta)=1-\frac{\theta}{\pi}$. This can be proven as follows. Consider two points on the sphere, separated by an angle $\theta\in \left[0,\pi\right]$. Assume a hemisphere colouring. Because half the sphere is coloured, the great circle through both these points will be cut into two equal halves: one coloured and one uncoloured. The probability this happens between the two points is $\frac{\theta}{\pi}$. Consequently, the probability these two points will be the same colour is $P_{hem}(\theta)=1-\frac{\theta}{\pi}$.

The Grasshopper Problem as defined above seeks bounds on the maximum success probability, but it also provides means for finding bounds on the minimum success probability. Because of the antipodality condition, there is a linear relation between the success probability of colouring $\mu$ for jumping angle $\theta \in \left[0,\pi\right]$ and angle $\pi-\theta$, given by $P_\mu(\pi-\theta)=1-P_\mu(\theta)$. This is understood as follows. If a point $A$ correlates with a point $B$ at angular separation $\theta$, for angle $\pi-\theta$ point $A$ will correlate with point $B'$, the point antipodal to $B$. Consequently, if $A$ and $B$ are the same colour this will constitute positively to $P_\mu(\theta)$, but because this also insinuates $B'$ is the opposite colour of $A$, it will constitute to $P_\mu(\pi-\theta)$ being lower. This can be generalised to the whole colouring $\mu$, giving $P_\mu(\pi-\theta)=1-P_\mu(\theta)$ for fixed $\mu$. This relation means using a set-up with jumping angle $\theta$ and maximising the success probability, should give a lower bound on the success probability for a set-up with jumping angle $\pi-\theta$. To summarise, this essay focuses on maximising the success probability for all $0\leq\theta\leq\pi$, and this simultaneously gives solutions for minimising the success probability.

In general we will need to turn to numerical techniques to solve the problem. The next section presents the numerical method.

\section{Numerical Set-up}
This essay researches the Grasshopper Problem numerically. To do so the problem has been discretised. The structure of the numerical approach is as follows. First, the sphere is discretised in an almost uniform antipodal\footnote{That is to say: for every point $\mathbf{r}$ there is a point $-\mathbf{r}$ and we call these points an antipodal pair.} grid. Secondly, for every pair of antipodal points exactly one is coloured, after which for every point the correlation with all coloured points is calculated (i.e. the probability the grasshopper will land on the grass after jumping up from a point). Finally, we try finding the optimal colouring for the probability function by searching through the space of colourings. For the last step two approaches are explored: a greedy algorithm and simulated annealing.

\subsection{Grid discretisation}
\label{sec:griddisc}
To perform numerical computations, the sphere's surface is discretised into an antipodal grid. A rectangular grid based on fixed longitudes and latitudes is unfavourable, since this leads to significant distortion of cell shape or area: square grids do not have equal area, whereas equal-area grids vary in shape from equator to poles. Instead in this essay a geodesic grid based on the icosahedron is used, known as a Goldberg polyhedron \cite{goldberg1937class}. This gives a grid which largely resembles a hexagonal grid. A geodesic grid leads to less distortion, as it overcomes oversampling at the poles and cells can be both minimally distorted in shape and area.

In this research an algorithm by Kurt von Laven \cite{Kurtgridsphere} is used to create the grid, which is based on the work of Nick A. Teanby \cite{teanby2006icosahedron}. This algorithm divides the triangular faces of the icosahedron into 4 equilateral triangles. This process is repeated $k-1$ times, after which the vertices are projected onto the unit sphere surface. The projection step induces deformations in the triangle size, which is especially apparent close to the original icosahedron's vertices.
Deformations can be reduced by using bubble meshing \cite{shimada1995bubble} or by shifting the vertices slightly to minimize area differences \cite{tegmark1996icosahedron}. This research does not make use of these methods. Parameter $k$ is referred to as the triangularisation depth. Define $N$ as the number of pairs of antipodal points, the total number of points $2N$ is given by: \cite{Kurtgridsphere}
\begin{equation}
    \label{eq:numpoints}
    2N=2 + 10(4^k).
\end{equation}
Additionally, $h$ is defined as a measure for the separation distance:
\begin{equation}
\label{eq:h}
    h=\sqrt{\frac{2\pi}{N}}.
\end{equation}
The projected vertices denote the centers of the cells. A vertex's cell is defined as the set of points on the unit sphere that is closest to the vertex. Because the grid of vertices is triangular, the cells are hexagonally shaped. The only exceptions are the 12 cells corresponding to the icosahedron's vertices, which are pentagons \cite{teanby2006icosahedron}. The icosahedron's symmetries ensure that the grid is antipodal. 

\begin{figure}
    \centering
    \includegraphics[width=1.0\textwidth,trim={0 0 0 2.15cm},clip]{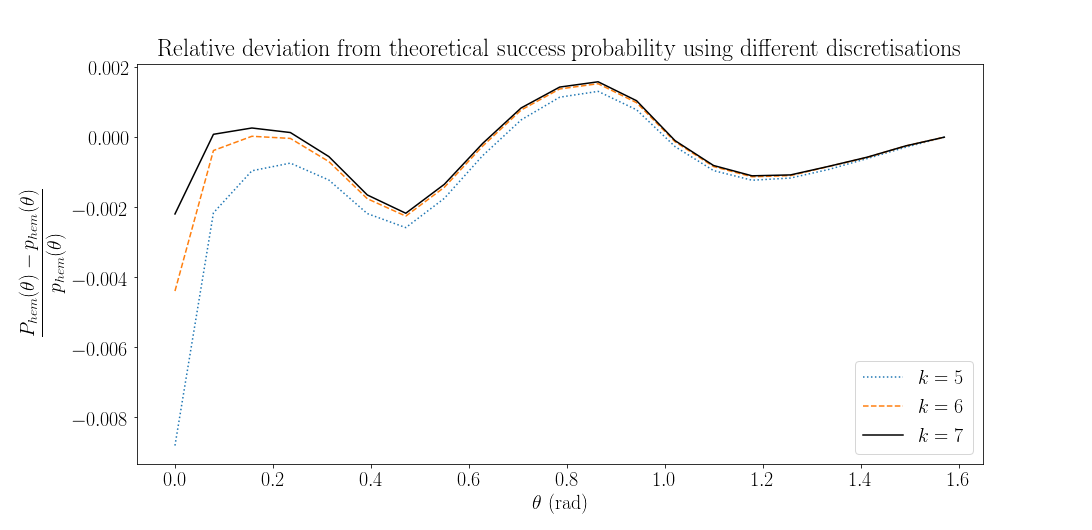}
    \caption{Deviation of success probability for a hemisphere colouring, relative to theoretical probability $p_{hem}(\theta)=1-\frac{\theta}{\pi}$ (see Sec. \ref{sec:premresults}), as a function of $\theta$ in the range $0\leq\theta\leq\frac{\pi}{2}$ and triangularisation depths $k=5$: $2N=10242$, $k=6$: $2N=40962$ and $k=7$: $2N=163842$.}
    \label{fig:discret_effects}
\end{figure}

In Figure \ref{fig:discret_effects} a comparison for different triangularisation depths is displayed for a hemisphere colouring.\footnote{There is a clear trend in the deviation, for which the reason is unknown. As this research uses $\theta>0.15$ and the deviation is marginal, this will not concern us.} For $\theta\geq0.1$ the results for $k=6$ and $k=7$ are within a $0.25\%$ margin of the theoretical value. Resolution $h$ scales with $2^{-k}$ and hence for small $\theta$ preference is given to $k=7$ ($h\approx0.009)$.

\subsection{Initialisation}
For initialisation $N$ points should be coloured such that the colouring is antipodal. For convenience let us call the $N$ points in the upper hemisphere by $U$,\footnote{Pairs on the interface can be split up arbitrarily} and the $N$ points antipodal to these by $L$. A colouring is now uniquely defined by $\mathbf s\in \{0,1\}^N$, where $s_i=1$ denotes vertex $i$ in $U$ to be coloured, $i$ in $D$ uncoloured, and $s_i=0$ vice versa. In a statistical physical perspective the discretised grid denotes a two-state spin system with every pair of antipodal points either spin up or down.\footnote{Equivalently one could define $\mathbf{s}\in \{-1,+1\}^N$} For initialisation three straightforward options are considered in this essay. 
\begin{itemize}
    \item $\mathbf s=\mathbf 1$ 
    \item $\mathbf s\in \{0,1\}^N$ randomly
    \item $\mathbf s$ equals the solution of a problem with a different jumping angle $\theta$.
\end{itemize}
In other words, the first option refers to a hemisphere colouring, the second to a random antipodal colouring and the third to using a colouring of previous calculations with a different jumping angle. In general this research will use a random initialisation, but the other initialisations can be used for verifying reliability of algorithms. See Sec. \ref{sec:othermethods} for comparisons.

\subsection{Correlation function}
Because the grid is discretised the delta function in the correlation function should be replaced by a smoothed approximation of the delta function. \cite{correlationfunction} explores the conditions a function like this should suffice. In this research the 4-point cosine function\cite{correlationfunction} is used:
\begin{align}
    \label{eq:corr}
    \phi\big(\frac{x}{h}\big)=&
    \begin{cases}
     \frac{1}{4}(1+\cos(\frac{\pi x}{2h}),\quad &\text{ if } \frac{|x|}{h}\leq 2\\
     0,\quad\quad\quad\quad\quad\quad &\text{ if } \frac{|x|}{h}>2
    \end{cases}
\end{align}
This function smears out the correlation over a $4h$ range. A different choice for the discretisation of the delta function can be made, of which a good analysis is given in \cite{yang2009smoothing}. \cite{goulkokent} compares this particular discrete delta function to a different choice in the context of the planar grasshopper problem. For the two tested functions this produces consistent results.

Let us now get an expression for the success probability of points in the discretised problem. Let $\mathbf{s}$ denote the colouring and $\theta$ the jumping angle. Given the discretised delta function, the correlation function between grid points $\mathbf v_i$ and $\mathbf v_j$ is of the form $\phi(\frac{\Delta\sigma(\mathbf v_i,\mathbf v_j)-\theta}{h})$, where $\Delta\sigma(\mathbf v_i,\mathbf v_j)$ is the central angle between the points. 
The success probability of an individual grid cell $i$ is the probability that a grasshopper, jumping up from grid cell $i$, lands on the lawn.\footnote{Note that this is also defined for points that are not part of the lawn.} It is defined as the total correlation between point $i$ and all points in the colouring, normalised by the total correlation with all points on the grid.\footnote{This guarantees that the success probability of every point remains between 0 and 1 even if there are irregularities in the grid. The secondary effect is that points will be normalised slightly differently, hence two correlated points will not be correlated to each other exactly the same amount. This will not concern us in further computations.} 
To ease notation, let $\mathbf{\tilde{s}}=\left[\mathbf{s}^\intercal, (\mathbf{1}-\mathbf{s})^{\intercal}\right]^\intercal$,\footnote{This means we double the length of vector $\mathbf{s}$, such that $ s_i=1\Leftrightarrow \tilde{s}_{i+N}=0$ and vice versa.} the success probability of point $\mathbf{v}_i$ is defined as:
\begin{equation}
    P_i=\frac{\sum_j \tilde s_j \phi(\frac{\Delta\sigma(\mathbf v_i,\mathbf v_j)-\theta}{h})}{\sum^{2N}_{j=1}\phi(\frac{\Delta\sigma(\mathbf v_i,\mathbf v_j)-\theta}{h})}.
\end{equation}
The success probability of a particular colouring $\mathbf{s}$ is trivially the average of the probabilities of points in the colouring:
\begin{equation}
     \label{eq:success}
    \begin{aligned}
    P_{\mathbf{s}}(\theta)&=\frac{1}{N} \sum_i \tilde{s}_i P_i\\
    &=\frac{1}{N}\sum_{i}\frac{\sum_{j}\tilde{s}_i \tilde{s}_j\phi(\frac{\Delta\sigma(\mathbf v_i,\mathbf v_j)-\theta}{h})}{\sum_j\phi(\frac{\Delta\sigma(\mathbf v_i,\mathbf v_j)-\theta}{h})}.
    \end{aligned}
\end{equation}
For calculating the correlation, there seems to be an arbitrariness in the choice for either angular distance $\psi=\Delta\sigma(\mathbf v_i,\mathbf v_j)$ or $l^2$ norm distance $d=||\mathbf v_i-\mathbf v_j||_2$, as a result of the bijection $d=2\sin(\frac{\psi}{2})$ for $0\leq \psi\leq \pi$. However, angular separation has as the clear advantage that it allows the use of a fixed $h$. To understand why this is the case, consider a circle formed out of equidistant points. Let us now inspect the correlation between points separated by angle $\psi$. If the correlation function is used as a function of the angular distance, for every angle a point will be correlated to approximately the same number of points. If, however, the correlation function is used with the $l^2$ norm distance $d$ as input and fixed $h$, a higher jumping distance would cause more points to fall in the band of correlated points; measuring from one specific point, points are not equidistantly separated as a function of $d$. This would also result in the antisymmetry breaking between set-ups with $\theta$ and $\pi-\theta$,\footnote{For any colouring and every $\theta$, theoretically $P_{\mathbf s}(\theta)+P_{\mathbf s}(\pi-\theta)=1$ as explained in Sec. \ref{sec:premresults}}: with $\theta<\frac{\pi}{2}$, the latter set-up would result in more correlations. As a fixed $h$ and fixed width of the correlation band is desirable, the use of angular separation is preferred over $l^2$ norm separation.

\subsection{Maximizing the probability}
As we have acquired an antipodal grid with $2N$ grid cells, the first $N$ to be antipodal to the second $N$ cells, the problem can be stated as finding an optimum over the hypercube $\{0,1\}^N$. For any reasonably accurate discretisation this space is huge. Two heuristic approaches are considered for trying to find a global maximum: a greedy algorithm and simulated annealing. Using both approaches might give information on the reliability of the methods, as well as insight into the optimization function itself.

\subsubsection{Greedy algorithm}
The first method is a greedy algorithm. The idea to use this method came from Zach Wissner-Gross's solution to the planar grasshopper problem \cite{greedyplan}. From a given colouring, the algorithm `flips' the pair that induces the highest increase in total probability. Flipping is defined as interchanging the colours of two antipodal points. This process is repeated until no such pair exists anymore. As a flip is only allowed if it increases the total probability, this method results in a monotonically increasing total probability. Consequently, this heuristic is prone to getting stuck in local optima. However, it could find a reasonably close approximation in relatively few steps. To partly circumvent this problem, multiple runs can be initiated using different initialisations. The difference in the resulting colourings is a measure of the algorithm's reliability. 

\subsubsection{Simulated annealing}\label{sec:SAmethod}
Simulated annealing \cite{kirkpatrick1983optimization} is tested as an alternative to the greedy algorithm. This algorithm is also used in \cite{goulkokent}, and therefore allows a better comparison to the planar results. Simulated annealing (SA) uses randomness to escape from local optima. Whereas a greedy algorithm always chooses the flip that induces the largest increase in success probability, SA allows flips that decrease the probability by less than a specified threshold. In this research an exponential cooling scheme is used. The procedure is as follows:
\begin{enumerate}
    \item Set initial temperature $T=T_0$, number of steps $m$ and $\alpha\in (0,1)$
    \item Choose random $i\in {1,...,N}$, $p$ from $U(0,1)$ 
    \item Calculate total success probability difference $\Delta P$ induced by flipping pair $i$
    \item Make flip $s_i:0\longleftrightarrow 1$ if Metropolis acceptance probability $\min(1,\exp{\frac{\Delta P}{T}})$ is larger than $p$, else pass
    \item Set $T$ to $\alpha T$ and repeat $m-1$ times from step 2
\end{enumerate}
Historically, simulated annealing was motivated by finding the lowest energy in physical systems, by gradually decreasing temperature-induced random effects. $T$ is therefore referred to as temperature, and the decreasing of temperature as cooling. The cooling rate should be slow enough, that the probability distribution of the current colouring is near the thermodynamic equilibrium at all times \cite{kirkpatrick1983optimization}. Note that in step 3 the flip is always accepted if this induces a positive change in the success probability. However, in contrast to the greedy algorithm there is also a chance this flip is accepted if it does not decrease the probability too much. Furthermore, a more negative $\Delta P$ (more decreasing total probability) and lower $T$ (a longer run time of the algorithm) leads to a lower acceptance ratio. When $T$ is very close to 0, a flip is only accepted if it hardly decreases the total success probability. This means at low $T$ almost all random points will not be accepted. Consequently it can be advantageous to use Monte Carlo updates \cite{newman1999monte} instead. This entails that in step 2 a pair is randomly selected from the subset of pairs whose probability of being accepted exceeds a set threshold. Subsequently the selected pair is always flipped. This method has not been used in this research. Noteworthy is that for small $T$ the algorithm does not become the described greedy algorithm, as the chosen flip does not necessarily correspond to the highest increase in probability.

Simulated annealing enables escape from local optima, but this comes at the cost of an additional complexity: a proper cooling algorithm should be chosen. An improper choice leads either to random flips for too long, or to a method that is unable to properly escape local optima. If the cooling starts off too slow, an ordered initialisation colouring will break down and become unordered. This is undesirable when a previous solution is used for initialisation. As the final colouring does not necessarily reflect that this might have happened, simulated annealing requires a close watch over the process itself. Additionally, more steps are required than in the greedy algorithm.

\section{Numerical Results}

In this section results for different settings are presented.\footnote{The author's Python implementation is available upon request} Three regimes emerge: a cogwheel regime $0.10\pi\leq\theta\leq 0.41\pi$, a critical regime $0.41\leq\theta\leq 0.55\pi$ and a cogs and stripes regime $0.55\pi\leq\theta\leq\pi$. Unless otherwise stated, results presented in this section are found using a grid with $2N=163842$ points,\footnote{Using Eq. \ref{eq:numpoints} with $k=7$} correlation function $\phi$ given by Eq. \ref{eq:corr}, a random antipodal colouring as initialisation and a simulated annealing approach.\footnote{Initial temperature $T_0=0.4$, $\alpha=0.99999$ and $150N\approx 10,000,000$ steps} Checks were performed for a different number of grid points ($2N=40962$), different initialisation methods and using instead the greedy algorithm. These alternative settings produce consistent results. At the end of this section results for different methods are compared.

\subsection{Cogwheel regime}
\begin{figure}
        \begin{subfigure}[b]{\textwidth}
            \includegraphics[width=\linewidth]{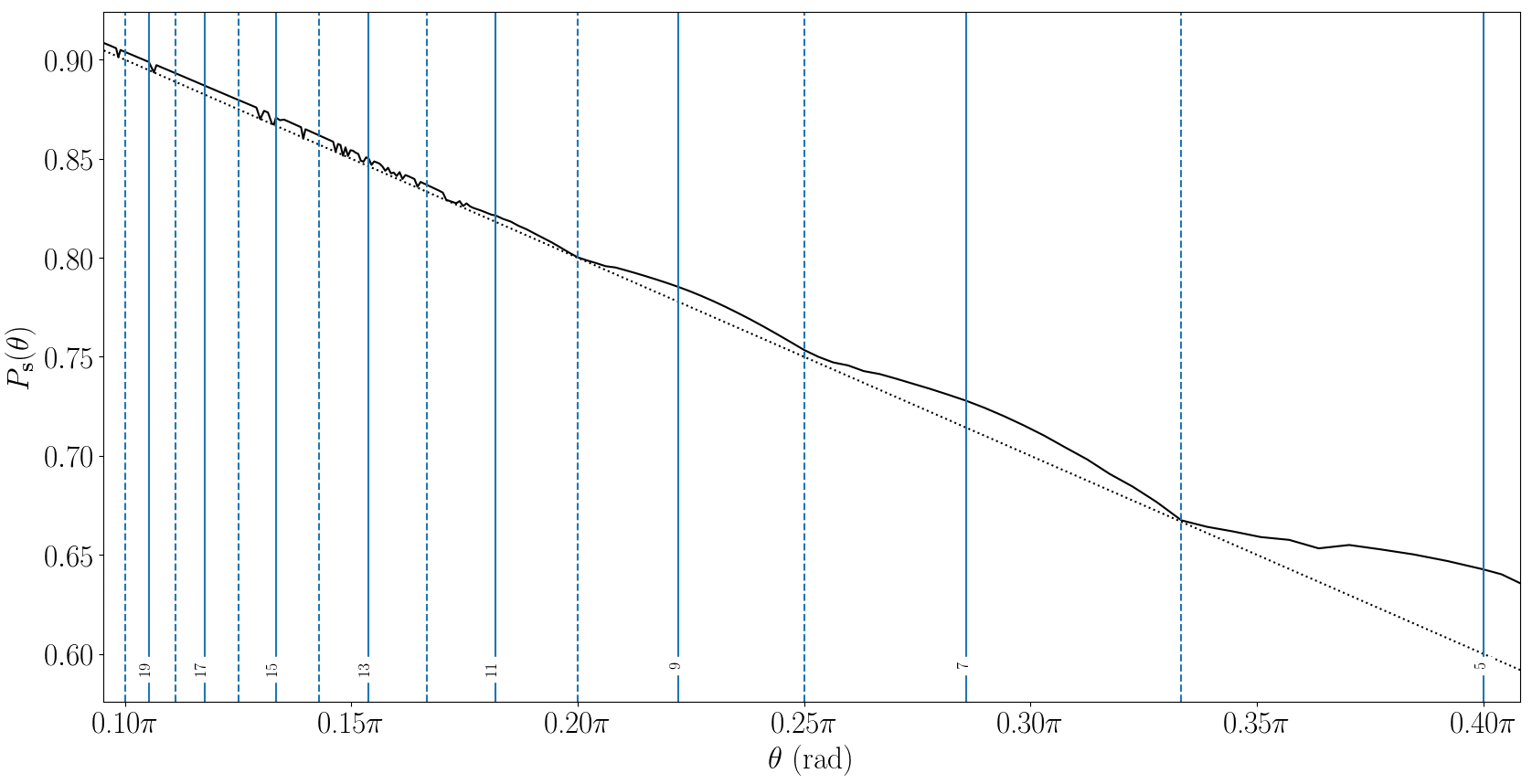}
            \subcaption{Success probability as a function of $\theta$ for generated colouring (solid line) and theoretical probability of hemisphere colouring (dotted).}
        \end{subfigure}%
        \hspace{\fill}
        \begin{subfigure}[b]{\textwidth}
            \includegraphics[width=\linewidth]{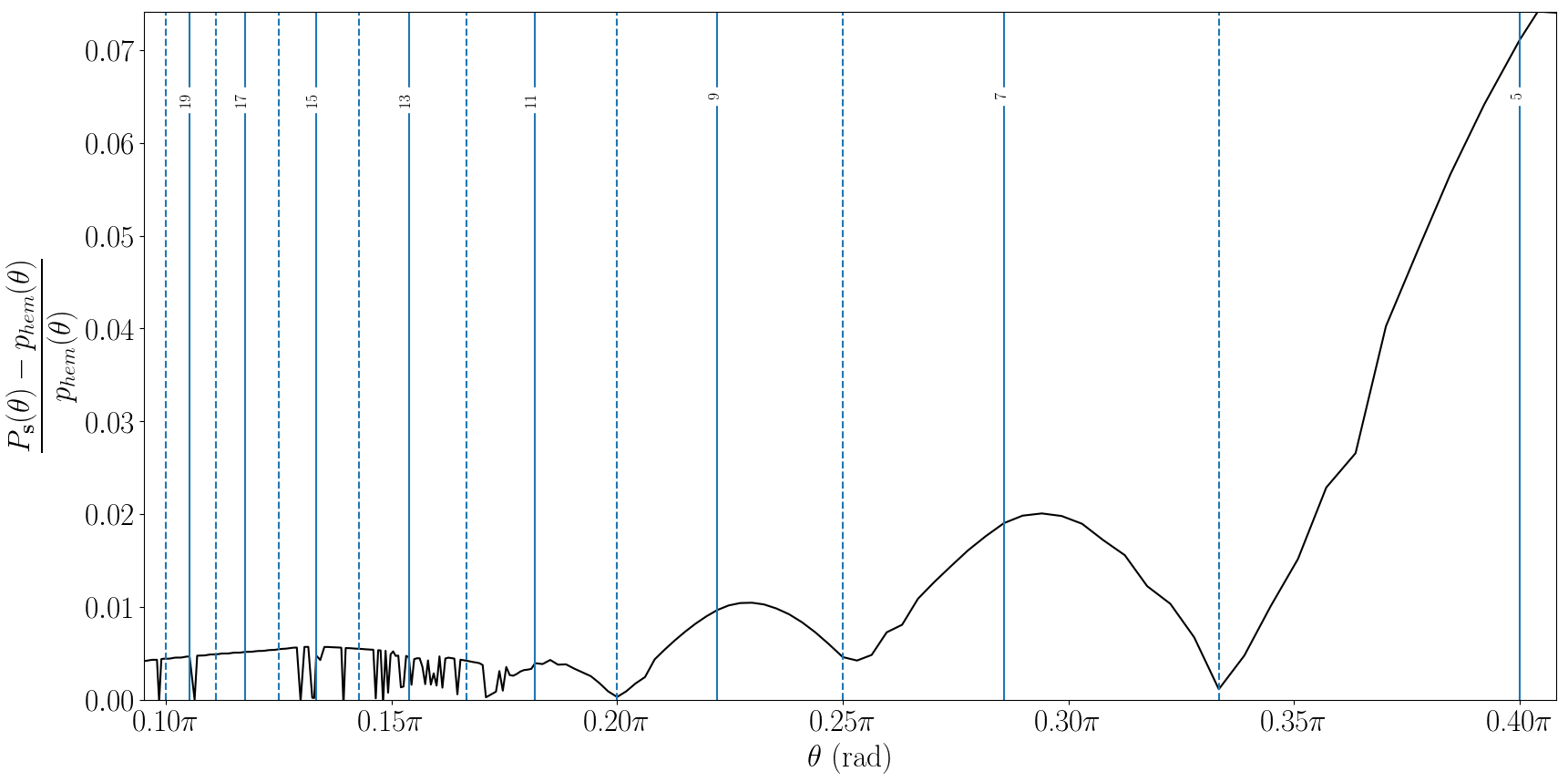}
            \subcaption{Success probability as a function of $\theta$, relative to hemisphere probability.}
        \end{subfigure}%
        \hspace{\fill}
        \caption{Success probabilities in critical regime $0.10\pi\leq\theta\leq 0.41\pi$, in comparison to theoretical hemisphere colouring success probability. The vertical lines denote values $\theta=\frac{2\pi}{n}$, $n$ integer}
        \label{fig:successcog}
\end{figure}

\begin{figure}
        \begin{subfigure}[b]{0.333\textwidth}
                \includegraphics[width=\linewidth]{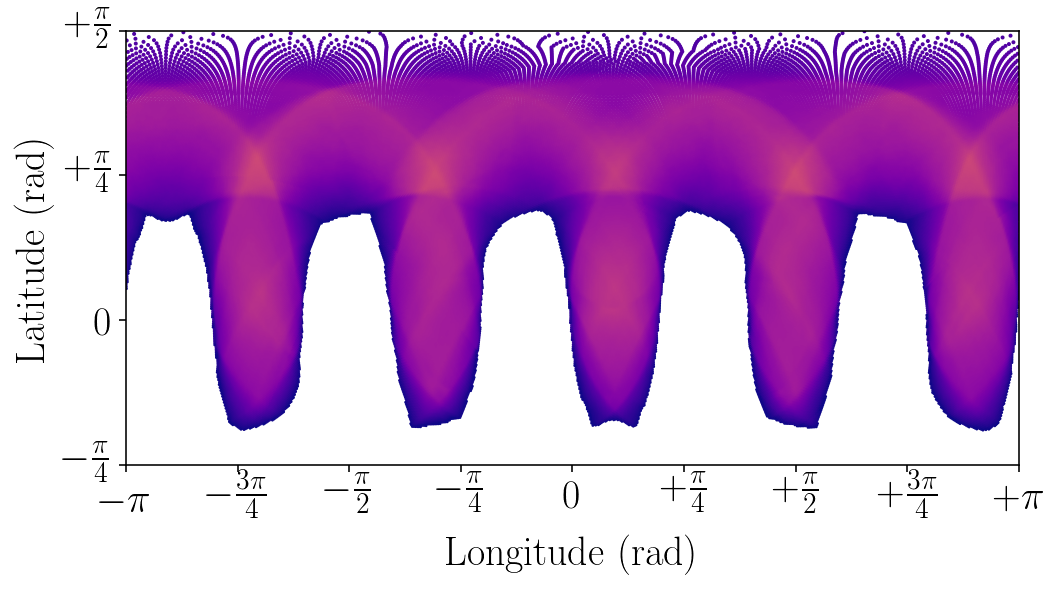}
                \subcaption{$\theta=\frac{2\pi}{5.0}= 0.4\pi$}
        \end{subfigure}%
        \begin{subfigure}[b]{0.333\textwidth}
                \includegraphics[width=\linewidth]{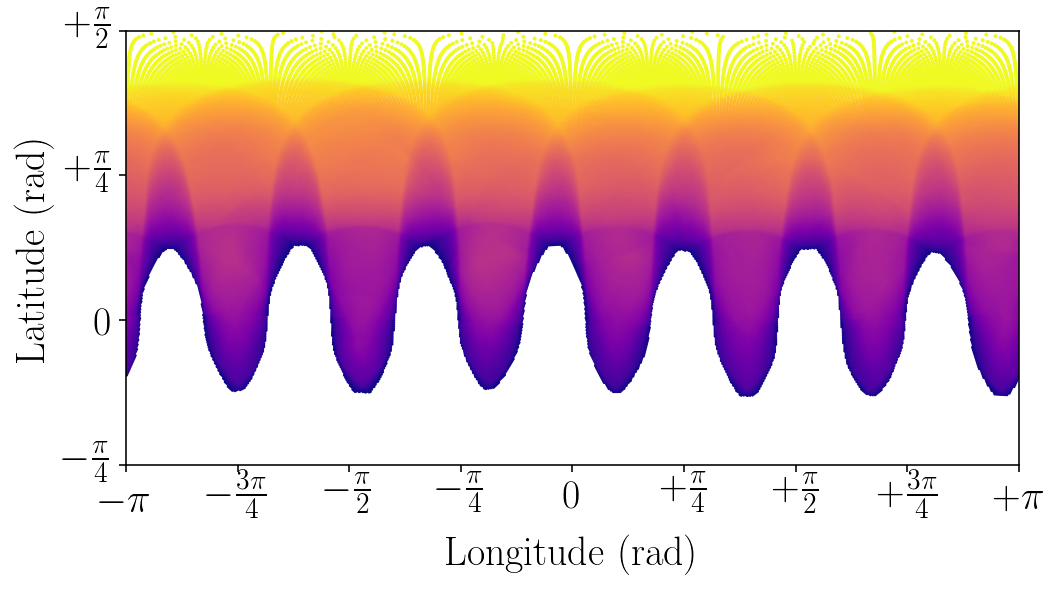} 
                \subcaption{$\theta=\frac{2\pi}{7.0}\approx 0.29\pi$}
        \end{subfigure}%
        \begin{subfigure}[b]{0.333\textwidth}
            \includegraphics[width=\linewidth]{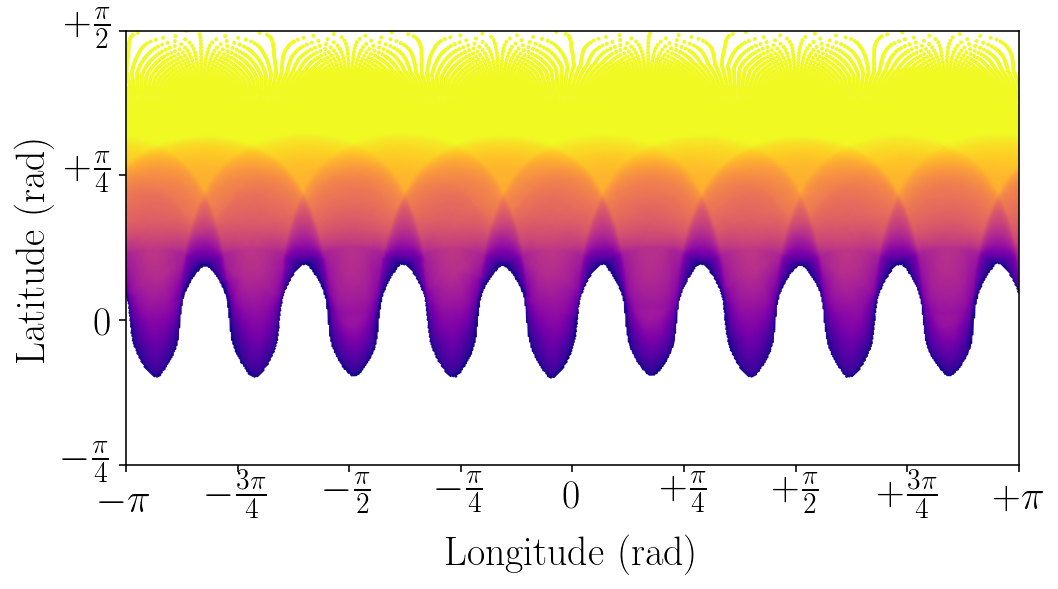}
        \subcaption{$\theta=\frac{2\pi}{9.0}\approx 0.22\pi$}
        \end{subfigure}%
        \hspace{\fill}
        \begin{subfigure}[b]{0.333\textwidth}
                \includegraphics[width=\linewidth]{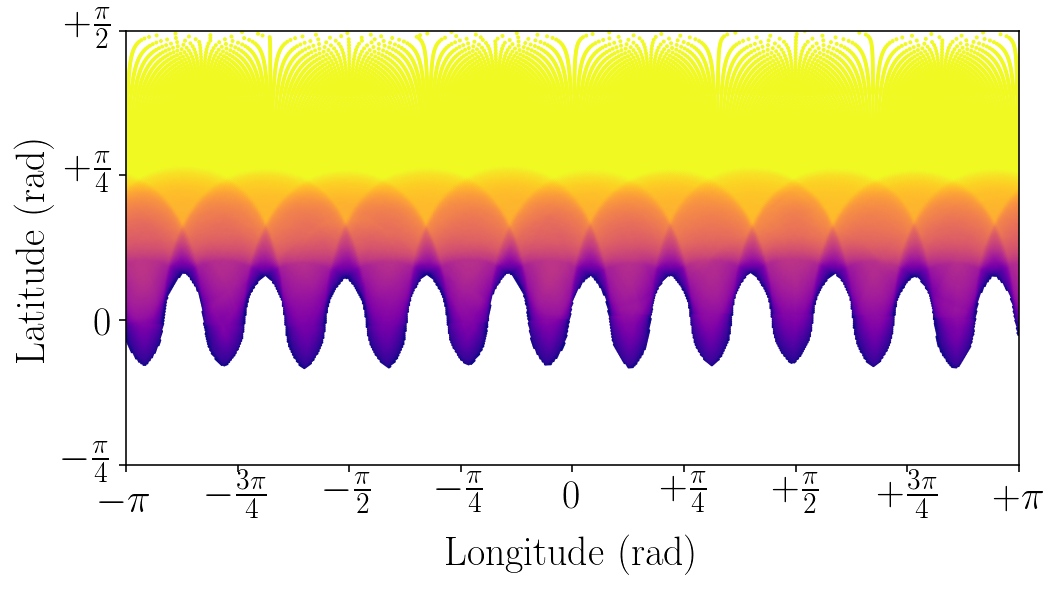}
            \subcaption{$\theta=\frac{2\pi}{11.0}\approx 0.18\pi$}
        \end{subfigure}%
        \begin{subfigure}[b]{0.333\textwidth}
                \includegraphics[width=\linewidth]{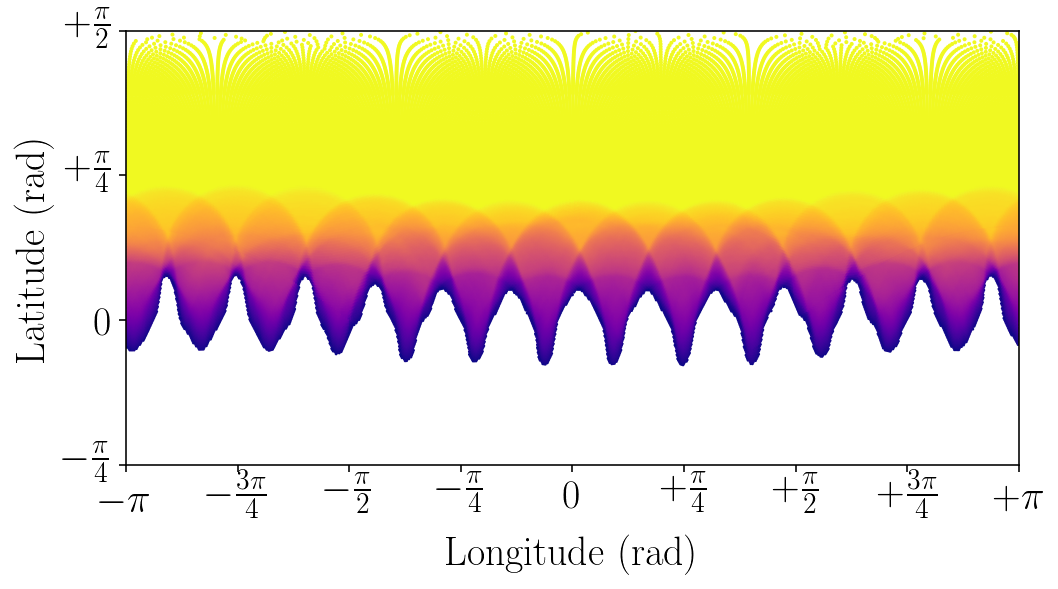}
        \subcaption{$\theta=\frac{2\pi}{13.0}\approx 0.15\pi$}
        \end{subfigure}%
        \begin{subfigure}[b]{0.333\textwidth}
                \includegraphics[width=\linewidth]{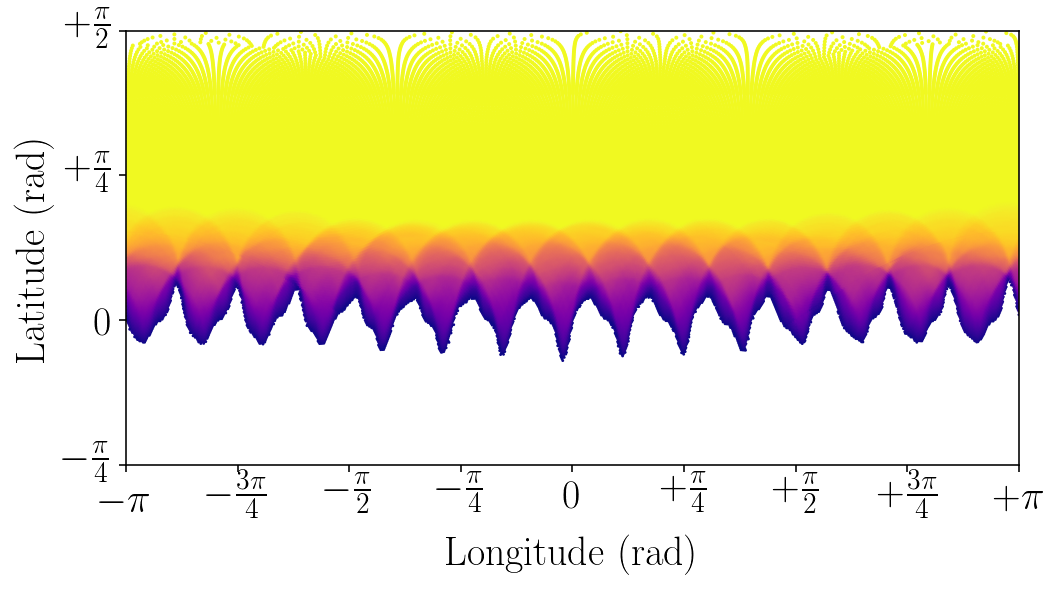}
                \subcaption{$\theta=\frac{2\pi}{15.0}\approx 0.13\pi$}
        \end{subfigure}%
        \hspace{\fill}
        \begin{subfigure}[b]{0.333\textwidth}
                \includegraphics[width=\linewidth]{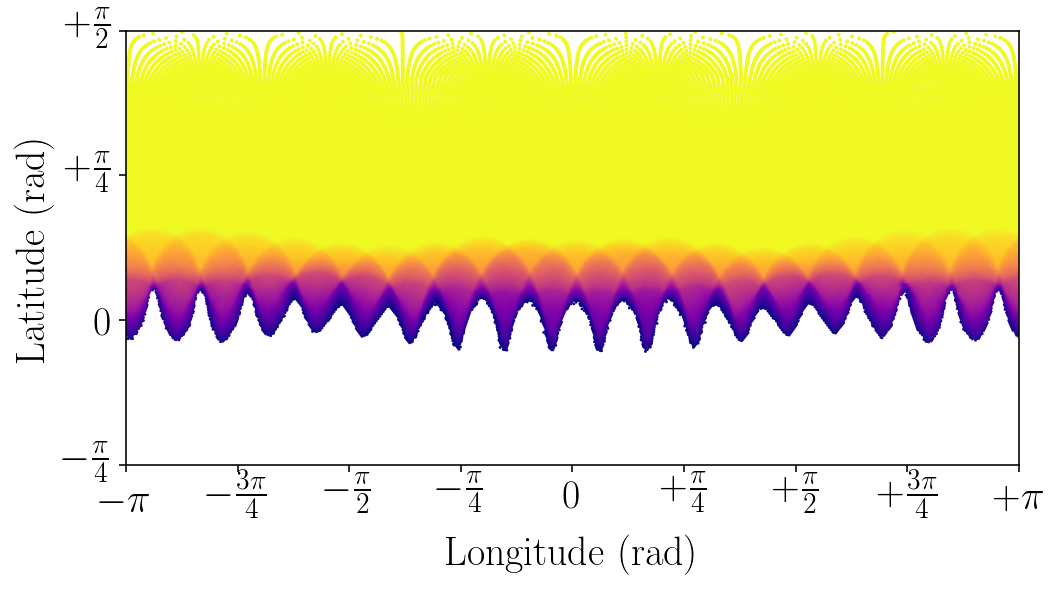}
                \subcaption{$\theta=\frac{2\pi}{19.0}\approx 0.11\pi$}
        \end{subfigure}%
        \begin{subfigure}[b]{0.333\textwidth}
                \includegraphics[width=\linewidth]{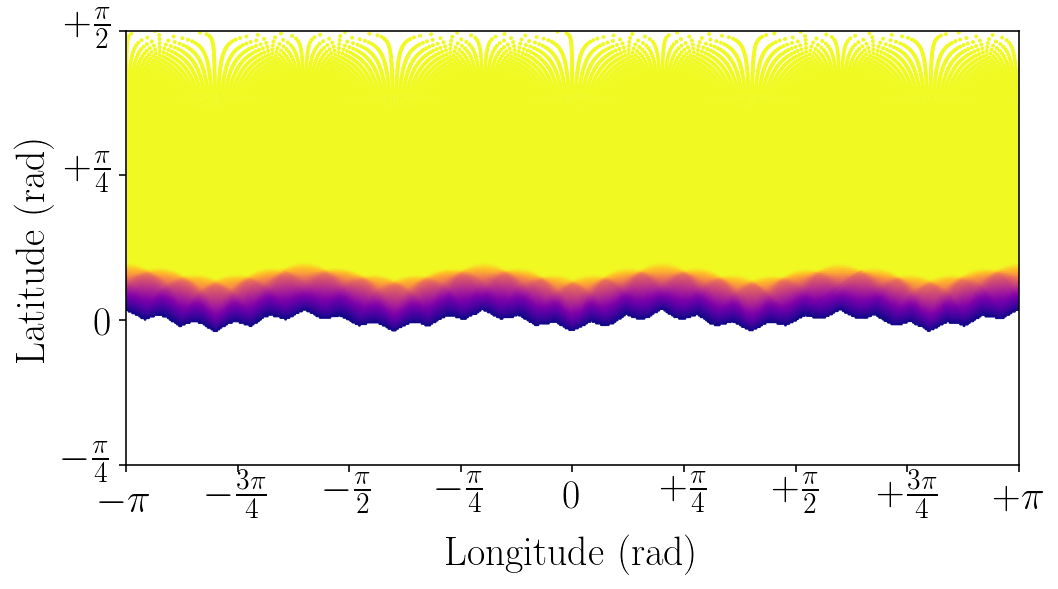}
                \subcaption{$\theta=\frac{2\pi}{25.0}=0.08\pi$}
        \end{subfigure}%
        \begin{subfigure}[b]{0.333\textwidth}
            \includegraphics[width=0.928\linewidth]{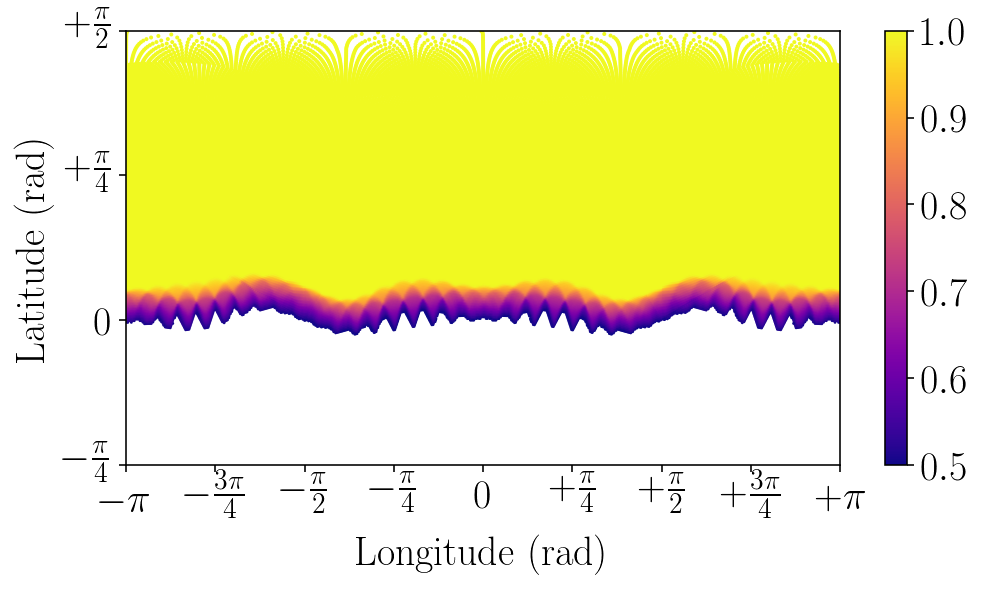}
            \subcaption{$\theta=\frac{2\pi}{37.0}\approx 0.05\pi$}
            \label{fig:37}
        \end{subfigure}%
        \hspace{\fill}
        \caption{Selection of cogwheel colourings for $\theta$ between $0.05\pi$ and $0.45\pi$. Colour denotes success probability of grid cells and the same colourmap is used for all figures. Colourings are rotated into northern hemisphere and the grid points are mapped on a longitude-latitude grid.}
        \label{fig:cogwheels}
\end{figure}

For $0.10\leq\theta\leq 0.41\pi$ colourings exhibit the cogwheel-like behaviour also found in \cite{goulkokent}. Figure \ref{fig:successcog} shows the success probability. For every angle, the success probability is higher than the theoretical success probability of the hemisphere colouring. In Figure \ref{fig:cogwheels} the optimal colourings for a range of jumping angles $\theta$ is shown. The quantity $c=\frac{2\pi}{\theta}$ is also noted, which in the case that $c$ is an odd integer\footnote{Because colourings are antipodal even number of equidistant cogs are impossible}, denotes the expected number of cogs. 

These colourings are in line with the planar result findings; cogwheel-shaped colourings appear to be optimal in case the angular distance between the cogs is approximately $\theta$. This preference for cogs spaced at intervals $\theta$ is reflected in Figure \ref{fig:successcog}. For angles smaller than $\frac{2\pi}{21}\approx 0.10\pi$ the system generally converges to a hemisphere colouring with irregular boundaries around the equator. As can be seen in Figure \ref{fig:37}, for some small angles cogs are found. Although the distance between the cogs is in this case indeed teh expected $\theta$, these colourings are generally irregularly shaped and only appear rarely. Additionally, in contrast to the discretisation errors as displayed in Fig. \ref{fig:discret_effects}, the relative success probability appears unreliable for low $\theta$. Possible explanations for the small $\theta$ discrepancies are given in the discussion.

When $\theta$ is far from $\frac{2\pi}{n_o}$, the success probability is significantly lower. 
Set-ups with angles sufficiently close to $\frac{2\pi}{n_o}$ with $n_o$ an odd integer, will generally converge to the closest `optimal' solution. Set-ups with values close to $\frac{2\pi}{n_e}$ with $n_e$ an even integer, will display more irregular behaviour. As can be seen in Figure \ref{fig:transitions}, this results in a lower success probability. The intermediate colourings of one of these transitions is studied further in Section \ref{sec:Transition}. Although these set-ups only converge badly, these still produce higher success probabilities than a hemisphere colouring (see Fig. \ref{fig:successcog}). Conclusively, for $\theta$ between $0.10\pi$ and $0.41\pi$ this essay's result give clear indication that cogwheel solutions are always more successful than hemisphere colourings.

If angle $\theta$ is a little larger than $\frac{2\pi}{n_o}$, this actually gives a higher success probability relative to the hemisphere solution. It is difficult to quantify this behaviour. An intuitive argument would be that a slightly larger angle does not change the success probability of cogs much, since the width of the cogs provides a buffer of the like. Consider a cogwheel colouring in which the northern hemisphere is mostly coloured. For points in the middle of the cogs the success probability when jumping longitudinally does not change with a slightly larger angle. The decrease in jumping probability is therefore mostly assigned to jumps with a large negative latitudinal component - jumping downwards from any point close enough to the edge of the cogwheel. However, the resulting loss in probability is similar for the hemisphere colouring, where the downwards component of jumps is the only component responsible for a jump's success. This means the change in success probability by increasing $\theta$ slightly from an optimal angle, is approximately equal for the hemisphere and cogwheel colouring. Because the hemisphere's success probability is lower for jumping angle $\frac{2\pi}{n_o}$ however, a further increase of jumping angle causes an increase in the cogwheel colouring's relative success probability. The success probability starts to decrease significantly when the success of mostly longitudinal jumps decreases even further; when $\theta$ starts to deviate too far from $\frac{2\pi}{n_o}$. 

\subsubsection{Transition}\label{sec:Transition}
\begin{figure}
        \begin{subfigure}[b]{0.333\textwidth}
                \includegraphics[width=\linewidth]{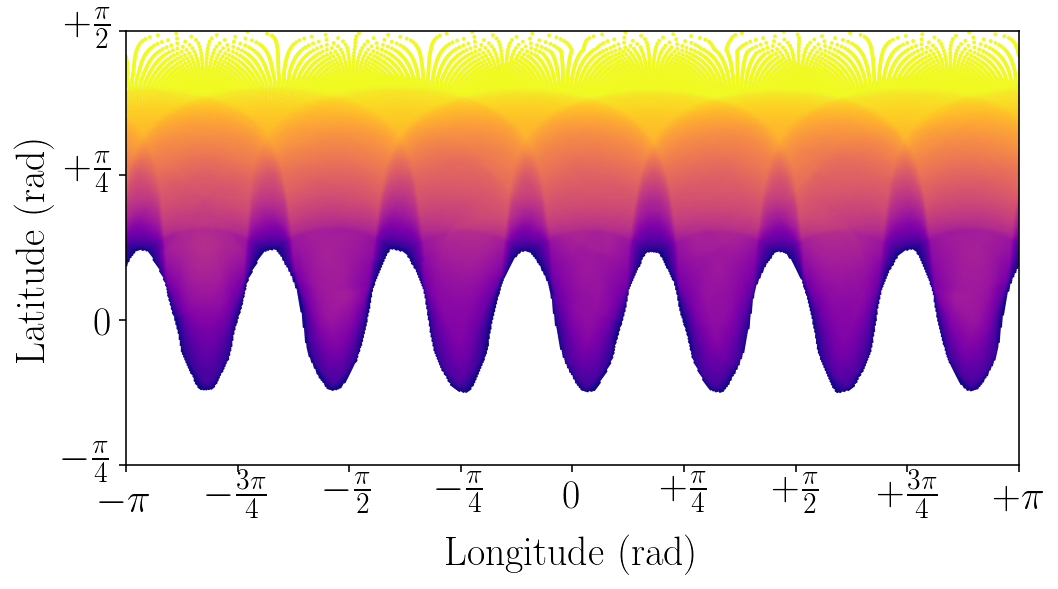}
                \subcaption{$\theta=\frac{2\pi}{7.2}\approx 0.278\pi$}
        \end{subfigure}%
        \begin{subfigure}[b]{0.333\textwidth}
                \includegraphics[width=\linewidth]{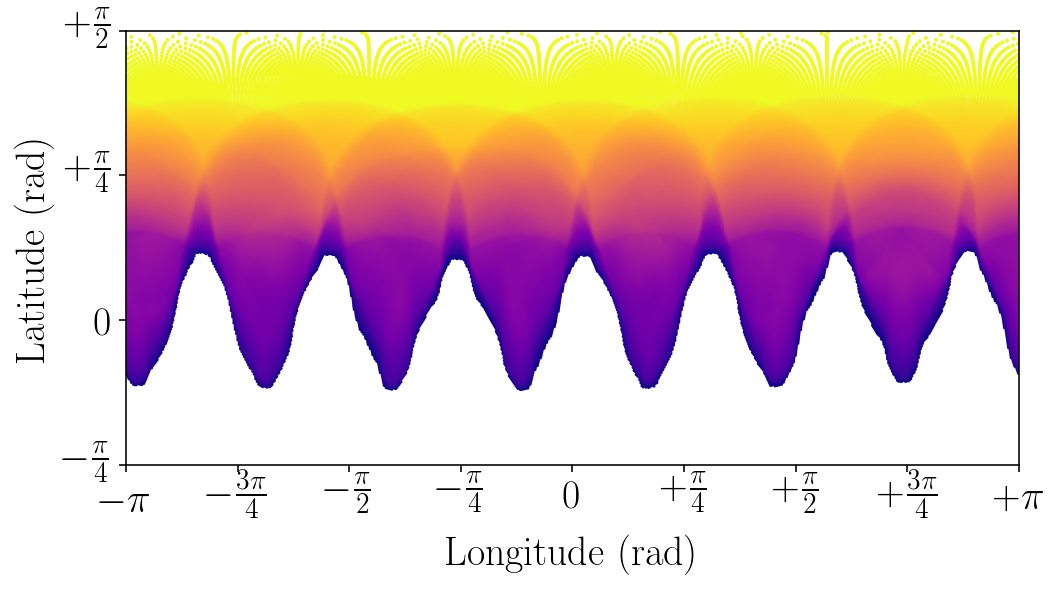}
             \subcaption{$\theta=\frac{2\pi}{7.5}\approx 0.267\pi$}
        \end{subfigure}%
        \begin{subfigure}[b]{0.333\textwidth}
                \includegraphics[width=\linewidth]{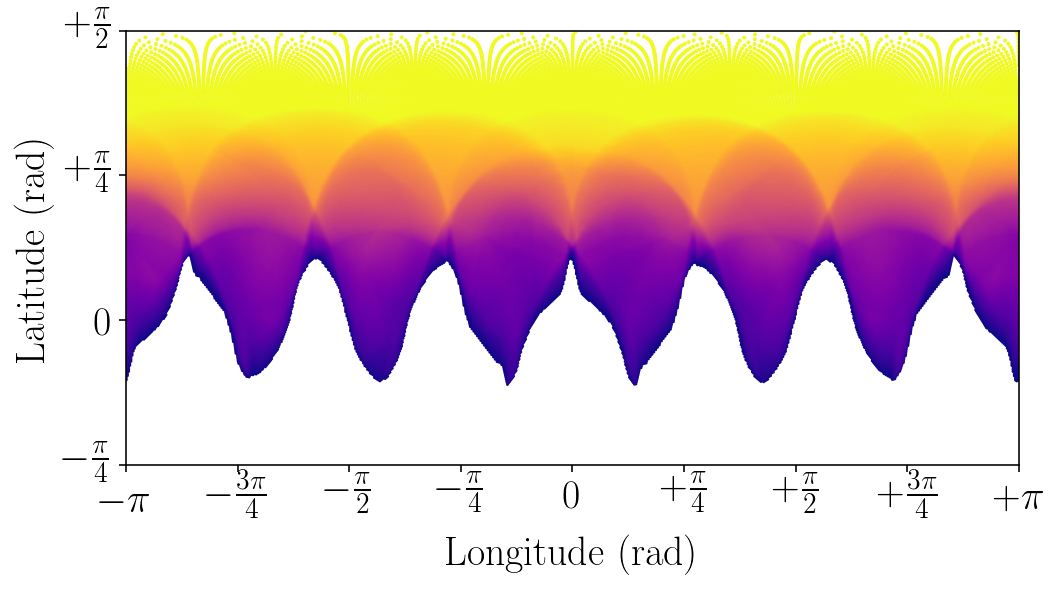}
                \subcaption{$\theta=\frac{2\pi}{7.8}\approx 0.256\pi$}
        \end{subfigure}%
        \hspace{\fill}
        
        \begin{subfigure}[b]{0.5\textwidth}
                \includegraphics[width=\linewidth]{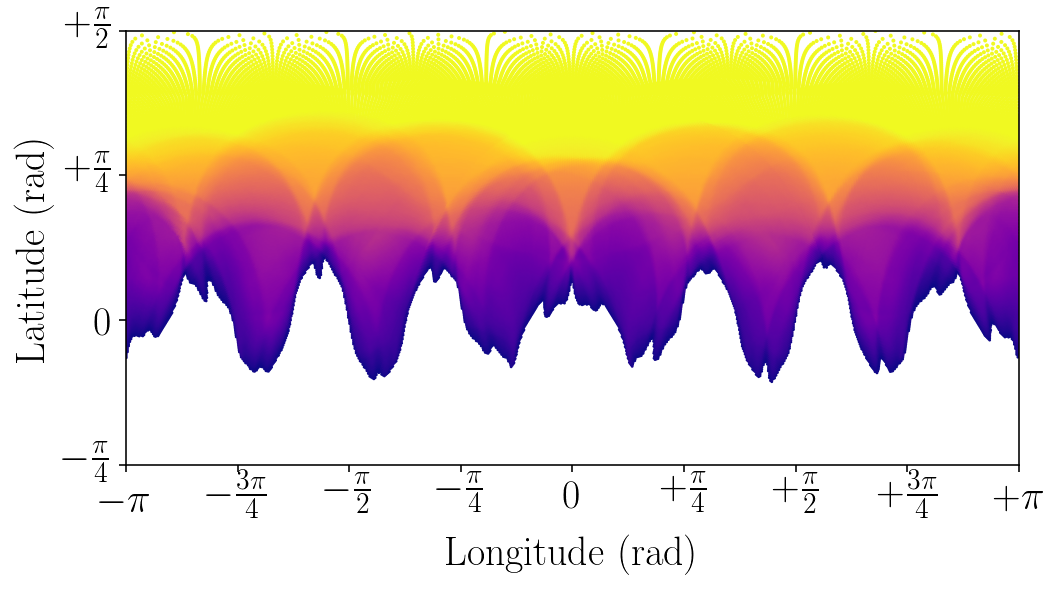}
                \subcaption{$\theta=\frac{2\pi}{7.9}\approx 0.253\pi$ (7-cog colouring)}
        \end{subfigure}%
        \begin{subfigure}[b]{0.5\textwidth}
                \includegraphics[width=\linewidth]{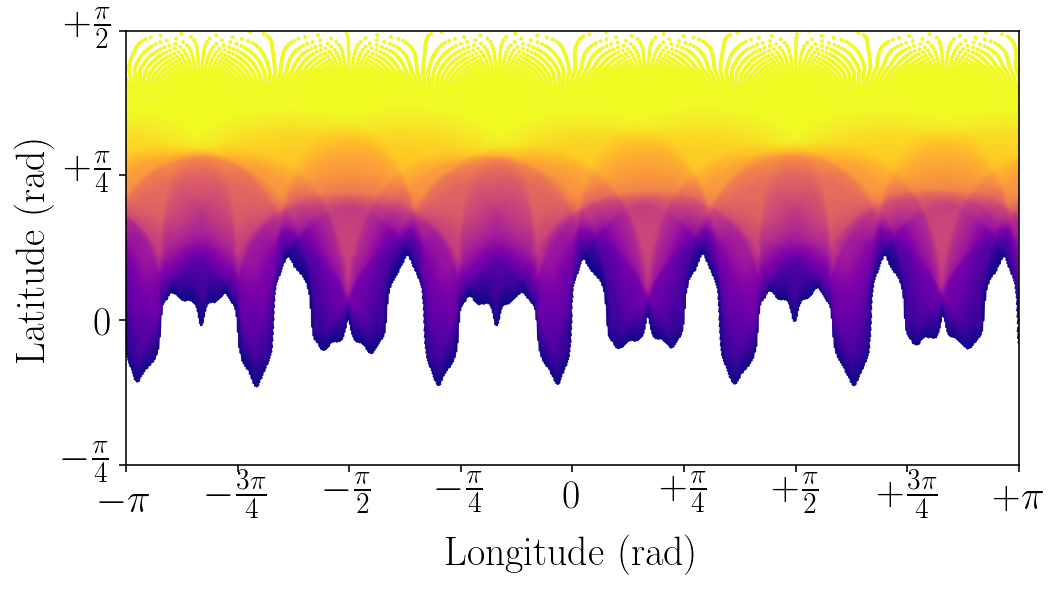}
             \subcaption{$\theta=\frac{2\pi}{7.9}\approx 0.253\pi$ (9-cog colouring)}
        \end{subfigure}%
        \hspace{\fill}
        
        \begin{subfigure}[b]{0.333\textwidth}
                \includegraphics[width=\linewidth]{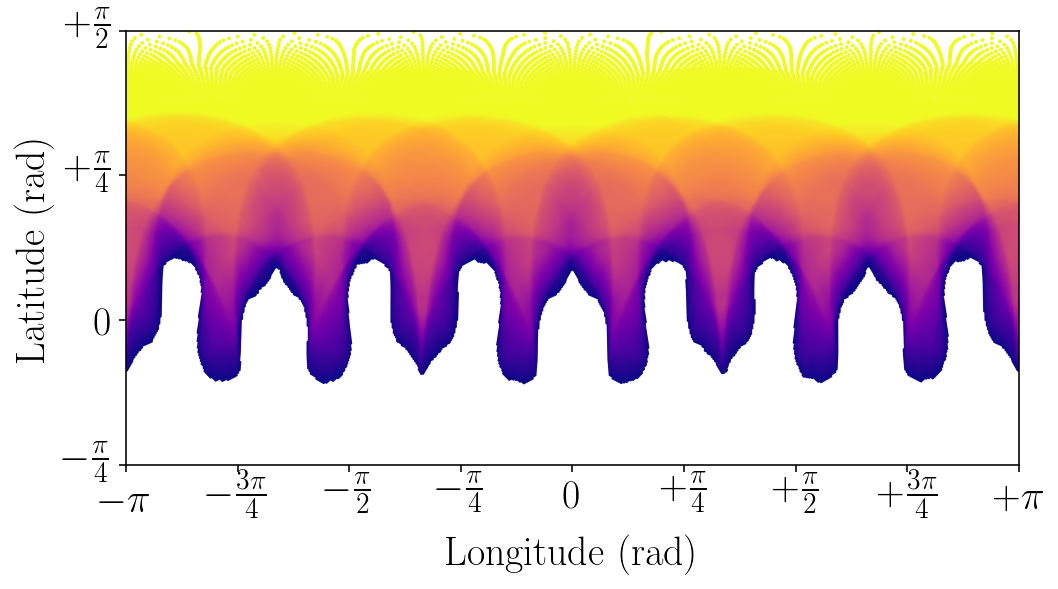}
                \subcaption{$\theta=\frac{2\pi}{8.0}=0.250\pi$}
        \end{subfigure}%
        \begin{subfigure}[b]{0.333\textwidth}
                \includegraphics[width=\linewidth]{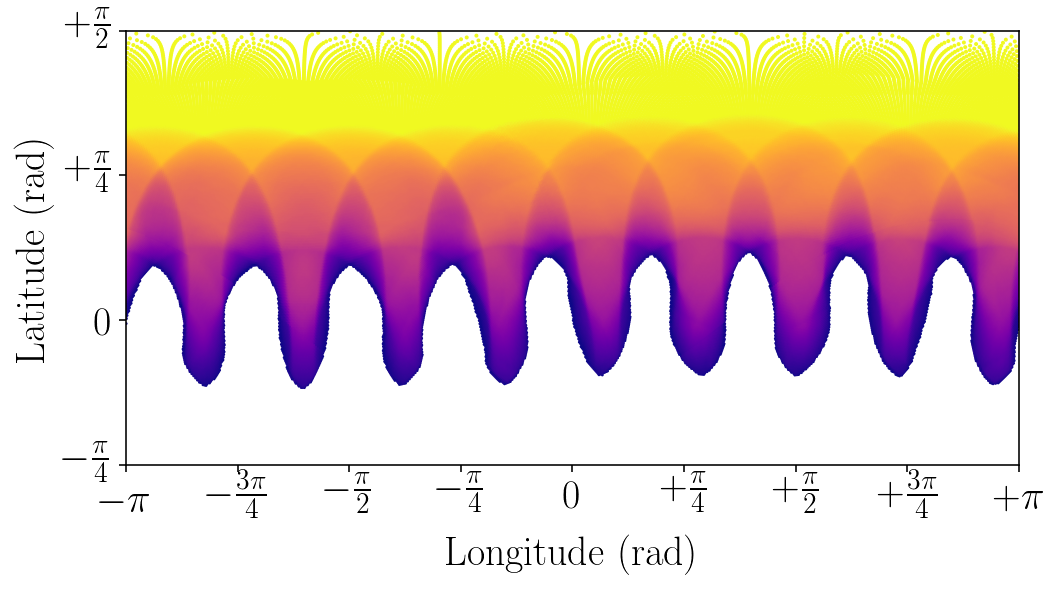}
             \subcaption{$\theta=\frac{2\pi}{8.4}\approx 0.238\pi$}
        \end{subfigure}%
        \begin{subfigure}[b]{0.333\textwidth}
                \includegraphics[width=0.928\linewidth]{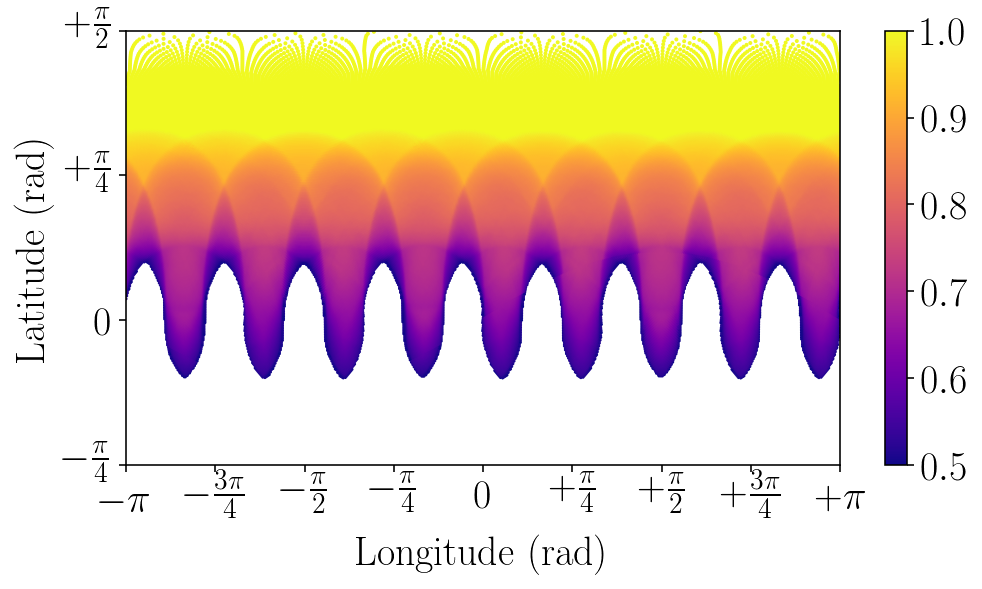}
                \subcaption{$\theta=\frac{2\pi}{8.8}\approx 0.227\pi$}
        \end{subfigure}%
        \caption{Colourings showing behaviour during transition from 7-cog cogwheel to 9-cog cogwheel. Colour denotes success probability of grid cells and the same colourmap is used for all figures.}        
        \label{fig:transitions}
\end{figure}

The different cogwheel solutions give convincing numerical proof that a hemisphere colouring is never optimal for $\theta=\frac{2\pi}{n_o}$, $n_o$ an odd integer. Interesting is the behaviour in the transition between these `optimal' solutions. Even using a slower cooling process,\footnote{$T_0=0.2$, $\alpha=0.999995$ and $150N$ number of steps} solutions often converge badly in between optimal angles, especially for angles $\theta<0.12\pi$. For the transition between $n_o=7$ and $n_o=9$ cogs, the most successful colourings out of five runs are displayed in Figure \ref{fig:transitions}, and success probability is shown in Figure \ref{fig:successcog}.

For $\frac{2\pi}{\theta}$ close to an odd integer, the most successful colouring has odd number of cogs closest to $\frac{2\pi}{\theta}$. In between these `optimal' colourings, the cogs look more disorderly. Different colourings show almost the same success probability at these angles, which results in an arbitrariness in the colouring the method produces. The most extreme example of this is shown for angle $\theta=\frac{2\pi}{7.9}\approx 0.253\pi$. The two colourings both have success probability 0.750, yet have a significantly different shape. Note that the number of major extrusions in these two colourings is nonetheless 7 and 9, as predicted.  

Multiple simulated annealing runs are used for each angle to find the above colourings. In some runs, methods get stuck in colourings with approximately double the predicted number of cogs. This can be explained as the cogs not correlating with their nearest cogs, but one cog further. \cite{goulkokent} predicts this behaviour, as colourings that exhibit the symmetries of a stellated polygon. These solutions have not been found to be more successful than the colouring with number of cogs equal to the odd integer closest to $\frac{2\pi}{\theta}$,\footnote{Although these solutions have been found if $\theta>\frac{\pi}{2}$, see Sec. \ref{subsec:stripes}} which is in agreement with the planar results.

\subsubsection{Comparison to planar results}
It is important to note that although the solutions look very similar to the cogwheel solutions of \cite{goulkokent}, there are three clear distinctions. First off, the colourings are truly antipodal. This results in coloured cogs being identically shaped as their antipodal uncoloured cogs. Secondly, this means the number of cogs is restricted to an odd number. Neither is the case for the planar results of \cite{goulkokent} as in the planar problem there was no intuitive way of defining antipodality. The cogs and the `holed out' parts are not alike in the planar case and even numbers of cogs are allowed. The difference in shape between holes and cogs in the planar case is further increased by the geometry of the disc: as the cogs are further from the disc's center than the holes, the distance between them is stretched compared to the holes. This is not the case for the spherical problem, as the cogs and holes are similarly placed around the equator, and thus are the same distance from the sphere's center. Thirdly, for transitions colourings are found that are different to the odd cog cogwheels. This is not consistent with the planar problem colourings, where systems always converged to a regular cogwheel. It is unknown whether this is the result of the problem's geometry, or a difference in the used methods. Further research into transitions is required.

\subsection{Critical regime}
\begin{figure}
        \begin{subfigure}[b]{0.5\textwidth}
                \includegraphics[width=\linewidth]{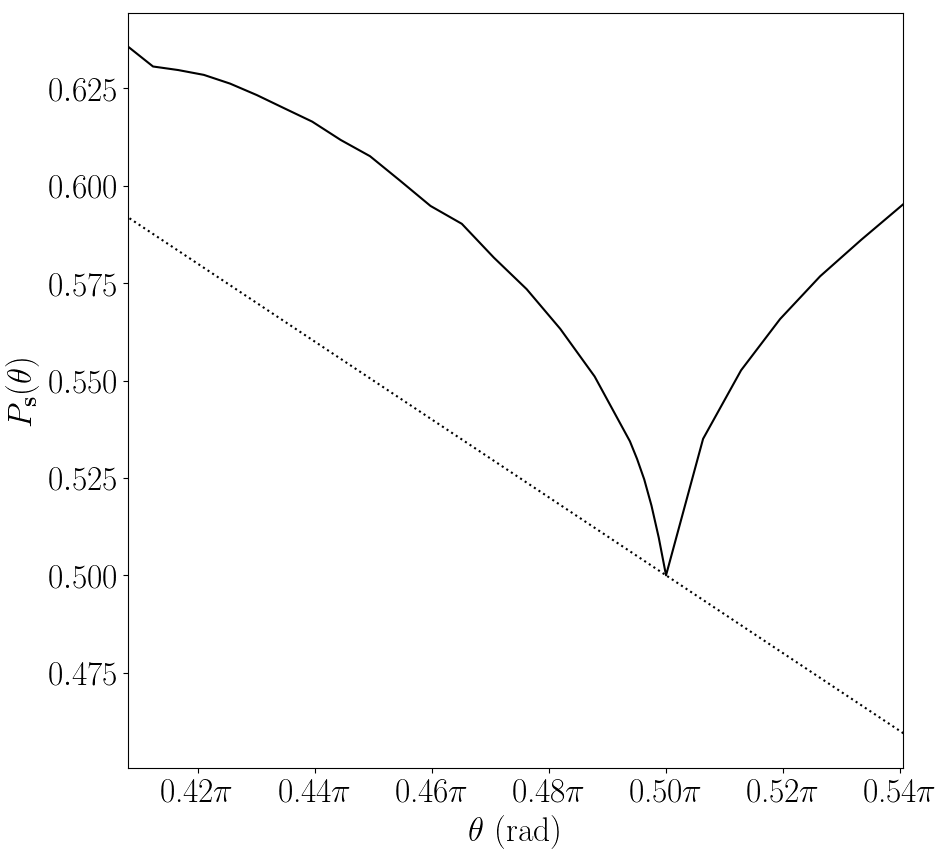}
                \subcaption{Absolute probability}
        \end{subfigure}%
        \begin{subfigure}[b]{0.5\textwidth}
                \includegraphics[width=\linewidth]{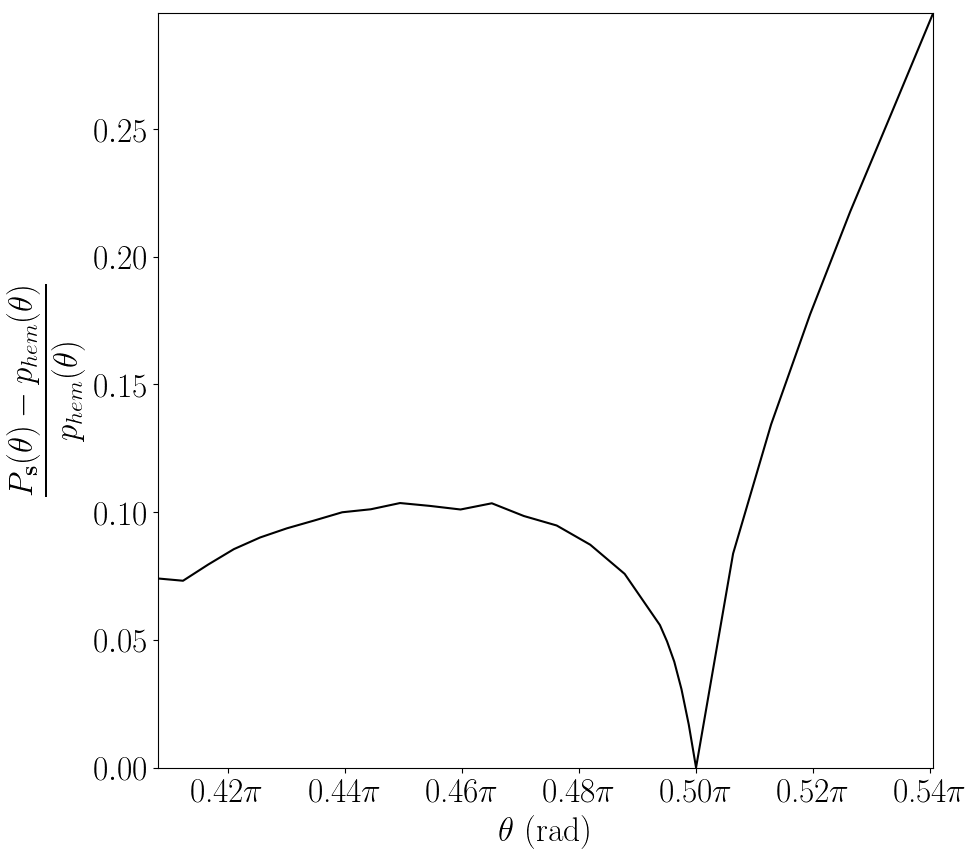}
                \subcaption{Relative probability}
        \end{subfigure}%
    \caption{Success probabilities in critical regime $0.41\pi\leq\theta\leq 0.55\pi$, in comparison to the theoretical success probability of the hemisphere colouring (drawn dotted in a)}\label{fig:successcritical}
    
\end{figure}

\begin{figure}
        \begin{subfigure}[b]{0.333\textwidth}
                \includegraphics[width=\linewidth]{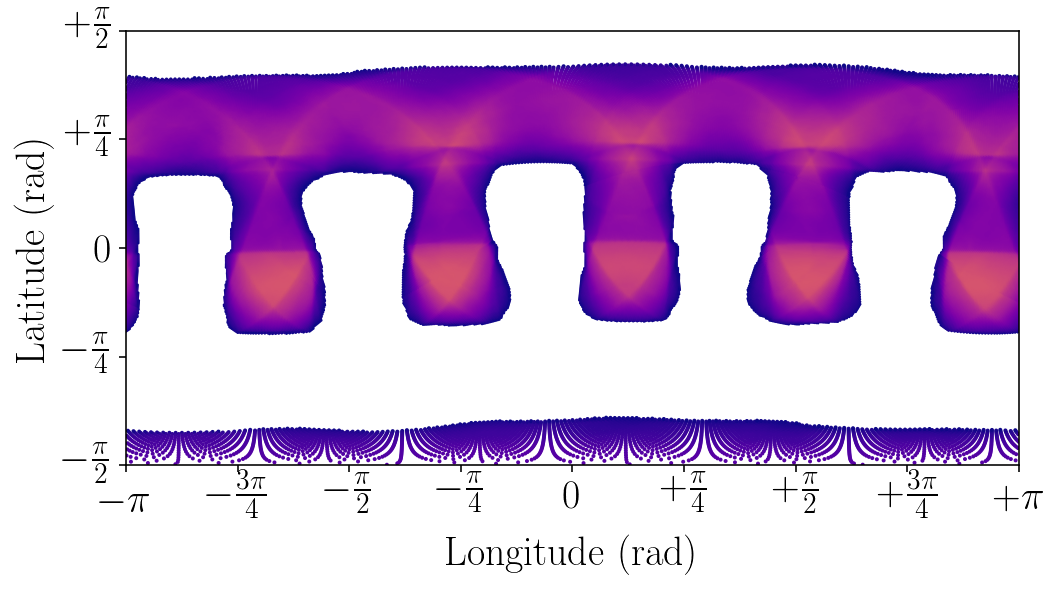}
                \subcaption{$\theta=\frac{2\pi}{4.90}\approx 0.408\pi$}
        \end{subfigure}%
        \begin{subfigure}[b]{0.333\textwidth}
                \includegraphics[width=\linewidth]{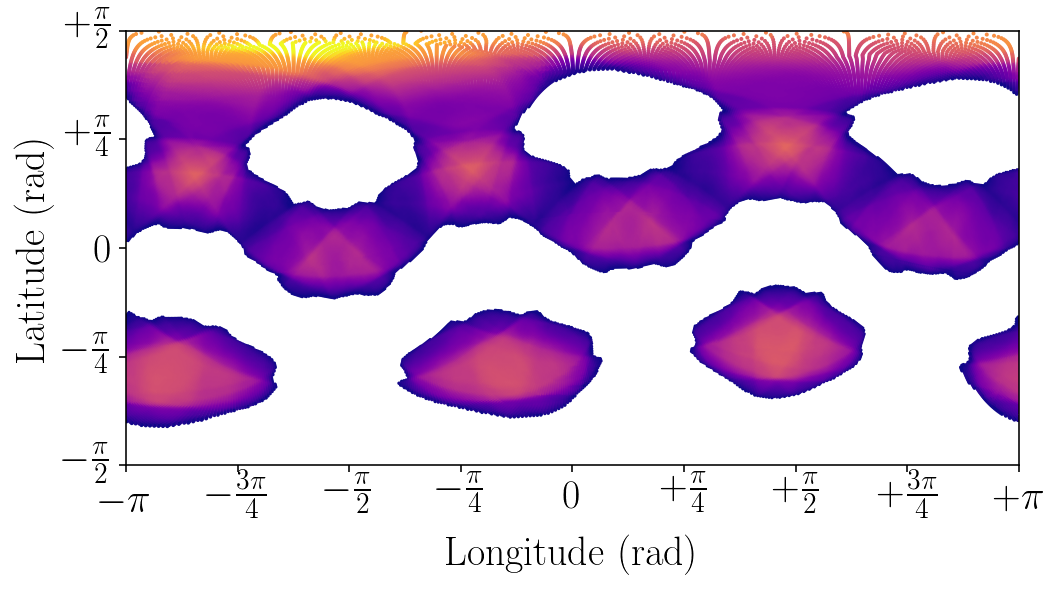}
                \subcaption{$\theta=\frac{2\pi}{4.85}\approx 0.412\pi$}
                \end{subfigure}%
        \begin{subfigure}[b]{0.333\textwidth}
                \includegraphics[width=\linewidth]{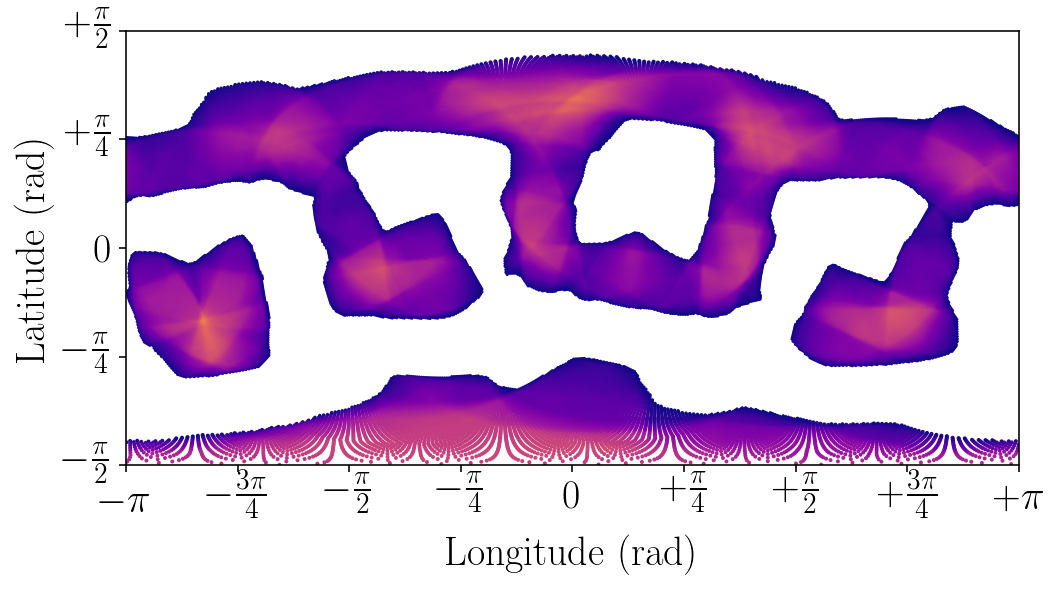}
                    \subcaption{$\theta=0.43\pi$}
        \end{subfigure}%
        \hspace{\fill}
        \begin{subfigure}[b]{0.333\textwidth}
                \includegraphics[width=\linewidth]{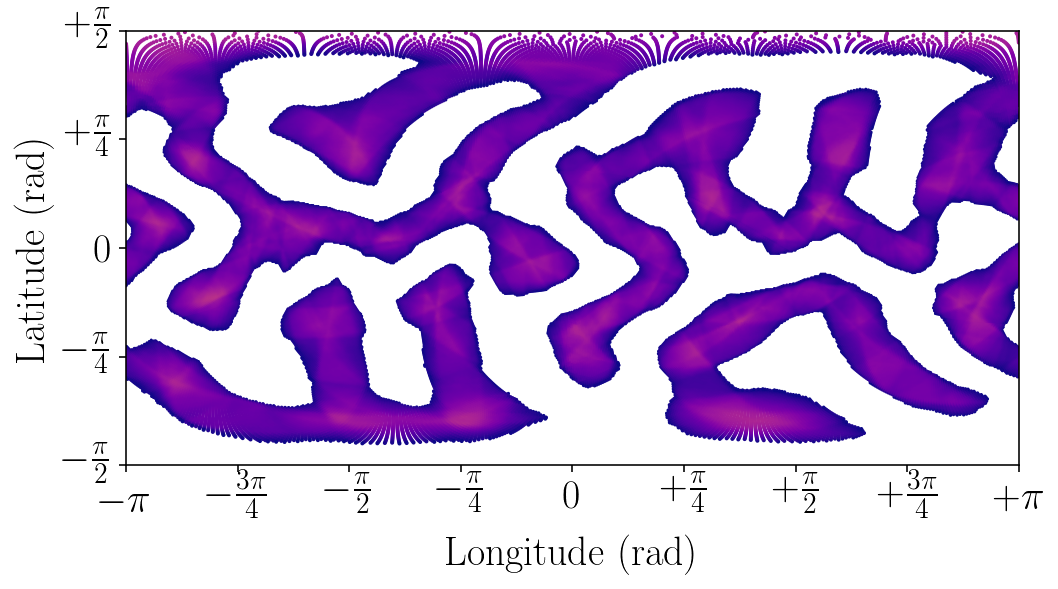}
        \subcaption{$\theta= 0.47\pi$}
        \end{subfigure}%
        \begin{subfigure}[b]{0.333\textwidth}
            \includegraphics[width=\linewidth]{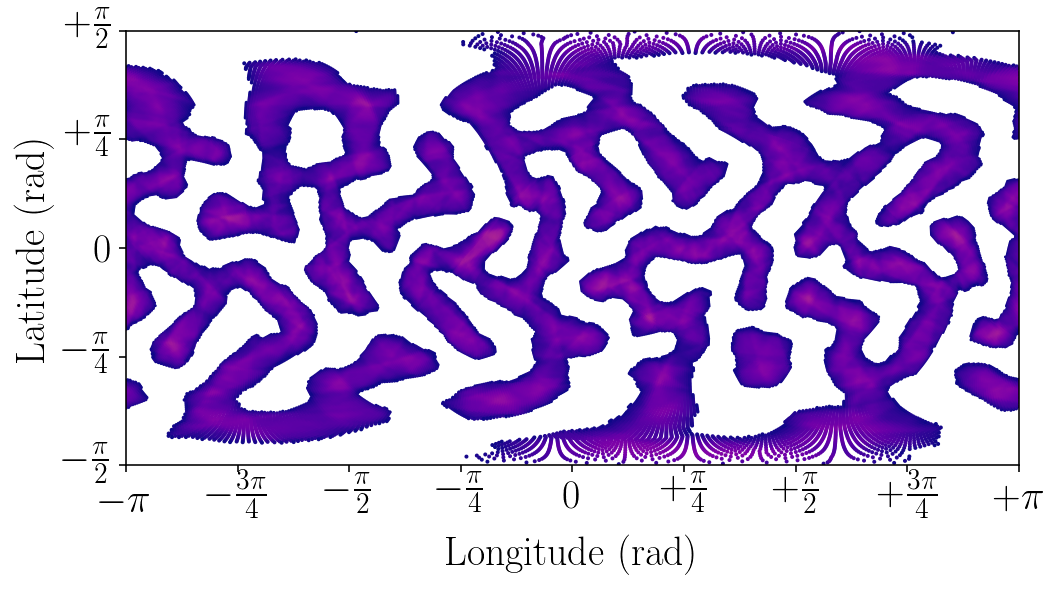}
            \subcaption{$\theta=0.48\pi$}
        \end{subfigure}%
        \begin{subfigure}[b]{0.333\textwidth}
                \includegraphics[width=\linewidth]{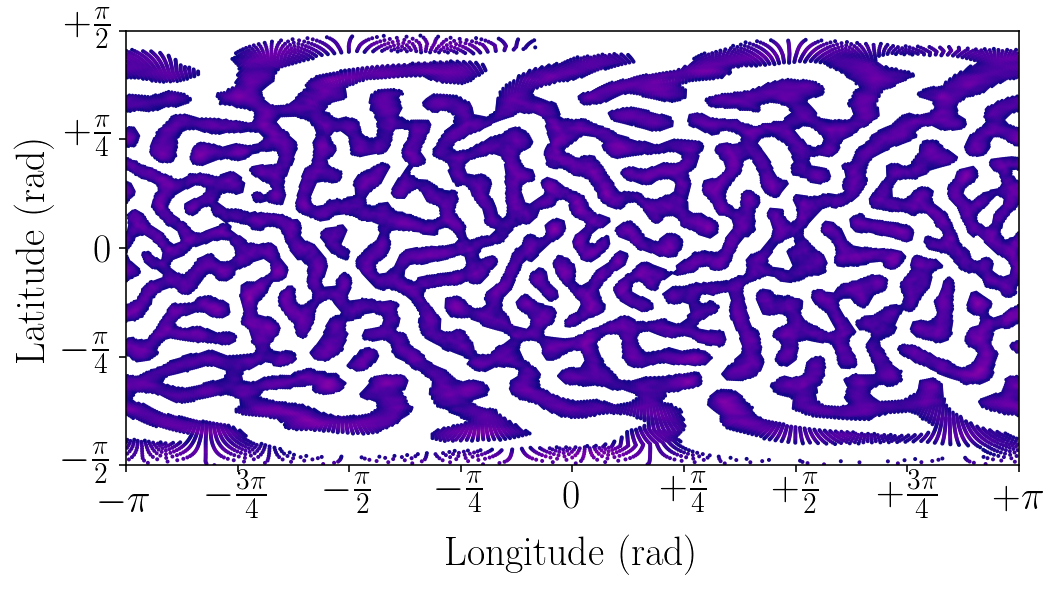}
        \subcaption{$\theta=0.488\pi$}
        \end{subfigure}%
        \hspace{\fill}
        \begin{subfigure}[b]{0.333\textwidth}
                \includegraphics[width=\linewidth]{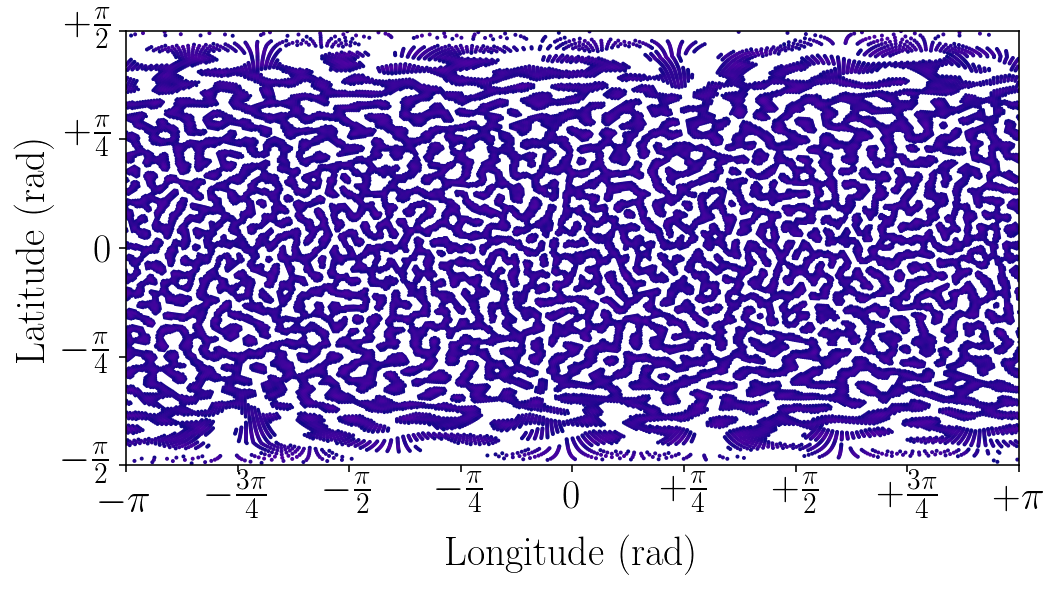}
        \subcaption{$\theta=0.495\pi$}
        \end{subfigure}%
        \begin{subfigure}[b]{0.333\textwidth}
                \includegraphics[width=\linewidth]{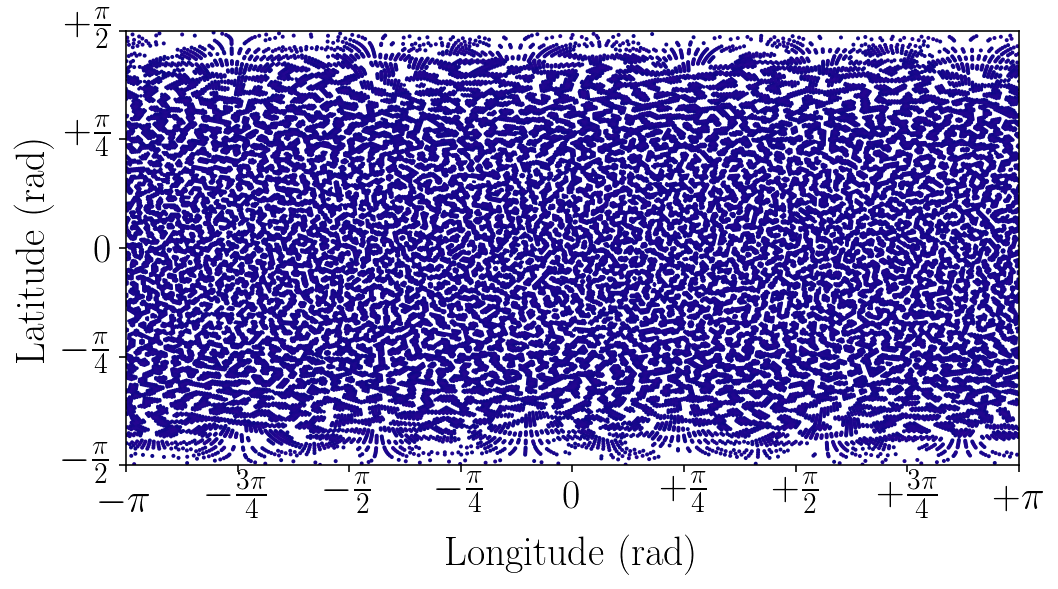}
        \subcaption{$\theta=0.499\pi$}
        \end{subfigure}%
                \begin{subfigure}[b]{0.333\textwidth}
                \includegraphics[width=\linewidth]{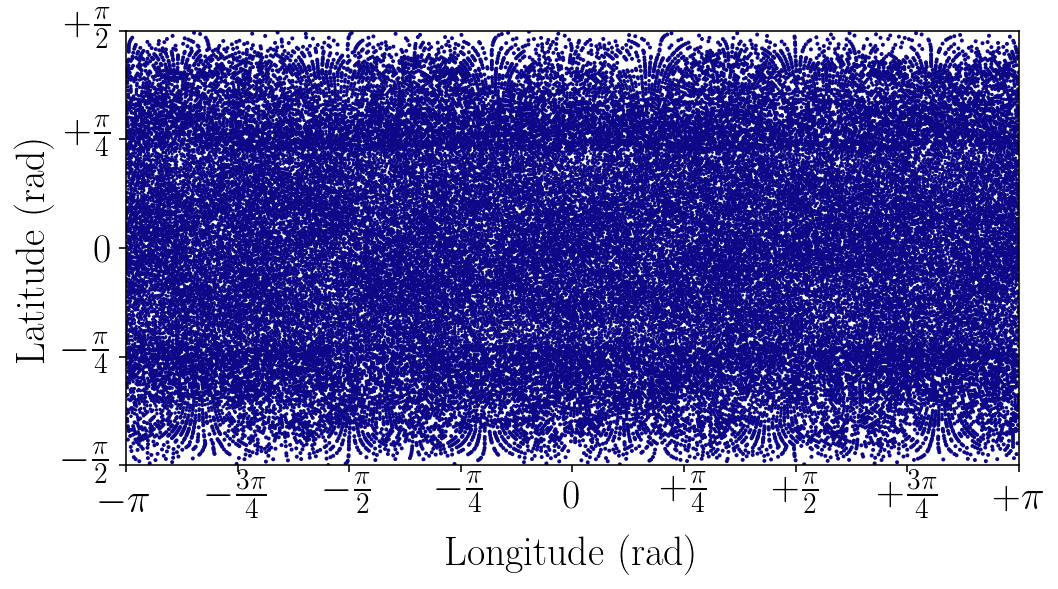}
        \subcaption{$\theta=\frac{\pi}{2}$}
        \end{subfigure}%
        \hspace{\fill}
        \begin{subfigure}[b]{0.333\textwidth}
                \includegraphics[width=\linewidth]{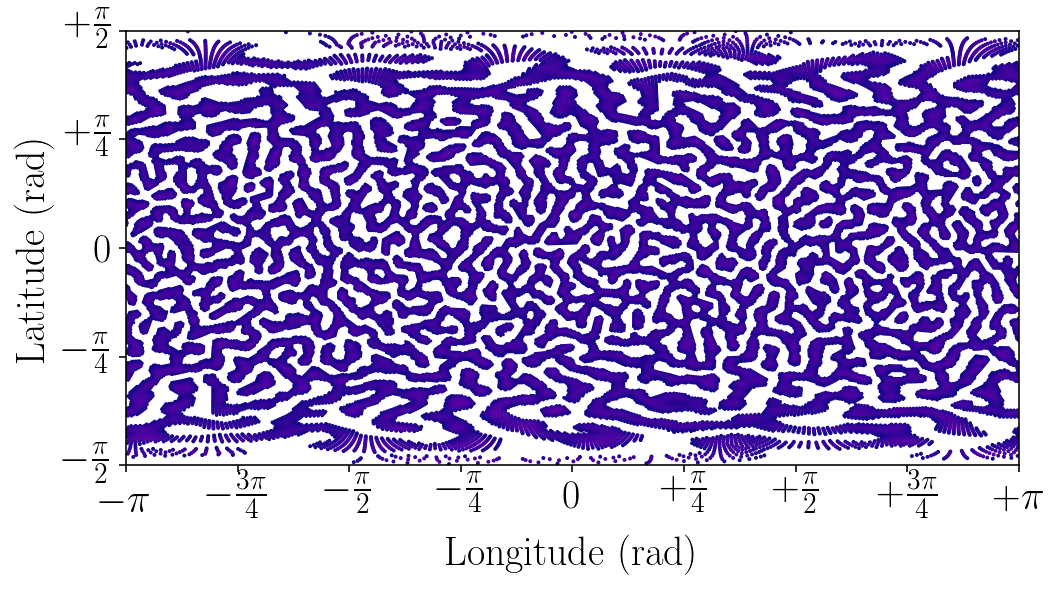}
        \subcaption{$\theta=0.506\pi$}
        \end{subfigure}%
        \begin{subfigure}[b]{0.333\textwidth}
                \includegraphics[width=\linewidth]{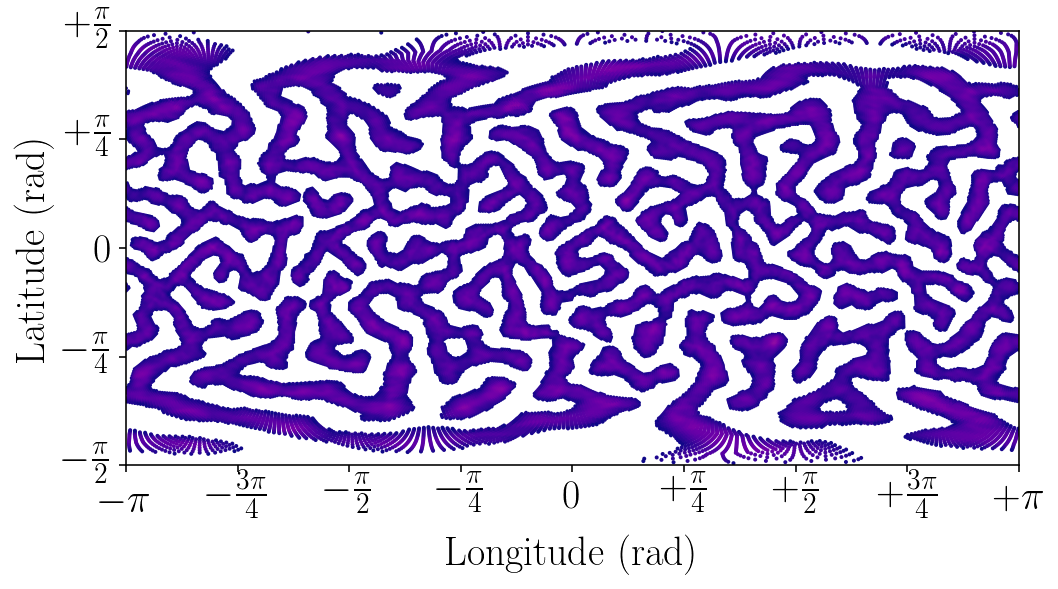}
        \subcaption{$\theta=0.512\pi$}
        \end{subfigure}%
        \begin{subfigure}[b]{0.333\textwidth}
                \includegraphics[width=0.928\linewidth]{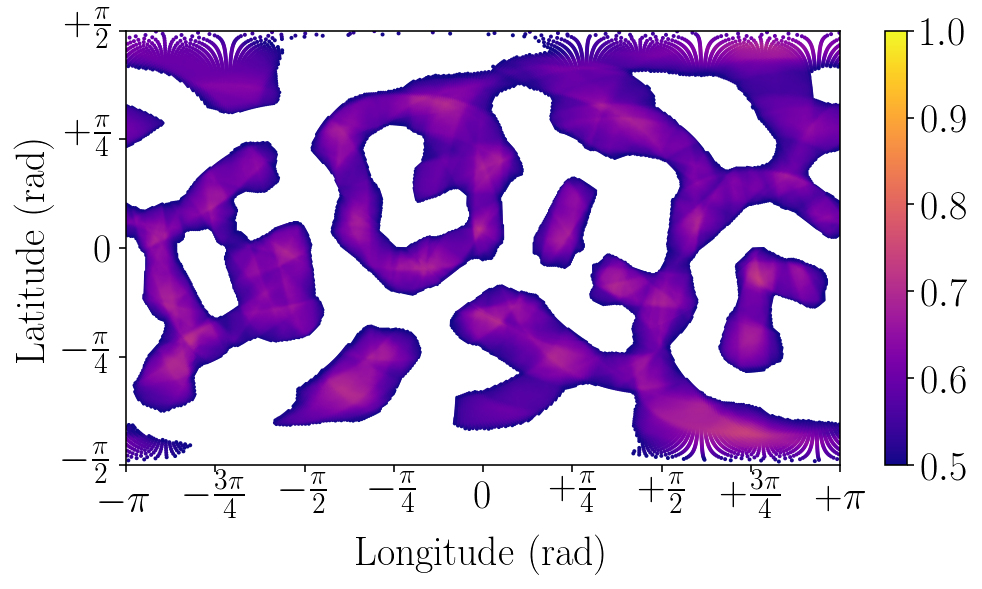}
        \subcaption{$\theta=0.541\pi$}
        \end{subfigure}%
        \caption{Colourings for $\theta$ between $0.41\pi$ and $0.55\pi$. Colour denotes success probability of grid cells and the same colourmap is used for all figures.}\label{fig:critical}
\end{figure}

In the regime $\theta\in\left[ 0.10\pi,0.41\pi\right]$ cogwheel solutions are found. If $\theta$ increases further, we do not get the same transition between a 5-cog colouring and a 3-cog colouring as seen earlier. Instead, more irregular colourings are generated. Figure \ref{fig:critical} shows the colourings as $\theta$ moves from $0.41\pi$ to $0.56\pi$. Figure \ref{fig:successcritical} shows the corresponding success probabilities.

The set-up converges to the 5-cogwheel solution if $\theta$ is smaller than $\frac{2\pi}{4.5}\approx 0.41\pi$. From this angle onwards, the colourings show chunks of coloured domains, which from $\theta\approx0.45\pi$ get smaller and smaller towards $\theta=\frac{\pi}{2}$. For $\theta>\frac{\pi}{2}$ the critical regime appears again and the same behaviour is observed: the coloured domains become larger until at $\theta= 0.55$ distinct shapes appear. Note that the absolute success probability is symmetrical around $\theta=0.5\pi$.

This behaviour is interesting as it resembles the behaviour of a simple Ising model with local interactions. When such a simple spin system is cooled down towards the critical point $T_c$, entropy decreases. As a result, domains will appear with equal spins. As the entropy decreases even further, these domains can become larger until the size becomes scaleless. This same behaviour is visible in the Grasshopper Problem colourings, but this happens as the angle $\theta$ moves away from $\frac{\pi}{2}$. 

The first difference is that in this essay's spherical system, the colouring with the most entropy is found for $\theta=\frac{\pi}{2}$ and the domains appear when moving away from this critical point. Another difference is the behaviour when moving closer to (further away from) the critical point for the simple infinite Ising lattice (the grasshopper sphere). In the former case, correlated domains will become unbounded in size, whereas in the grasshopper case the cogwheel solution becomes visible. To summarise, the spherical grasshopper system displays behaviour similar to a simple Ising model on an infinite lattice, as close to jumping angle $\theta=\frac{\pi}{2}$ it acts as if it is ruled by a reduced temperature of the form $\frac{1}{|\pi-2\theta|}$. This analogy breaks down for $\theta\approx (0.50\pm0.05)\pi$.

Notable is the behaviour at exactly $\frac{\pi}{2}$. For $\theta=\frac{\pi}{2}$ any colouring should give a probability of $\frac{1}{2}$, as it correlates with antipodal pairs of points. This means flipping the colouring of a pair of points leads to no difference in the success probability. A simulated annealing method will consequently always accept a flip, hence the resulting colouring is merely the result of the pseudorandom nature of the simulated annealing algorithm.

\subsection{Cogs and stripes regime}\label{subsec:stripes}
\begin{figure}
    \centering
    \includegraphics[width=\linewidth]{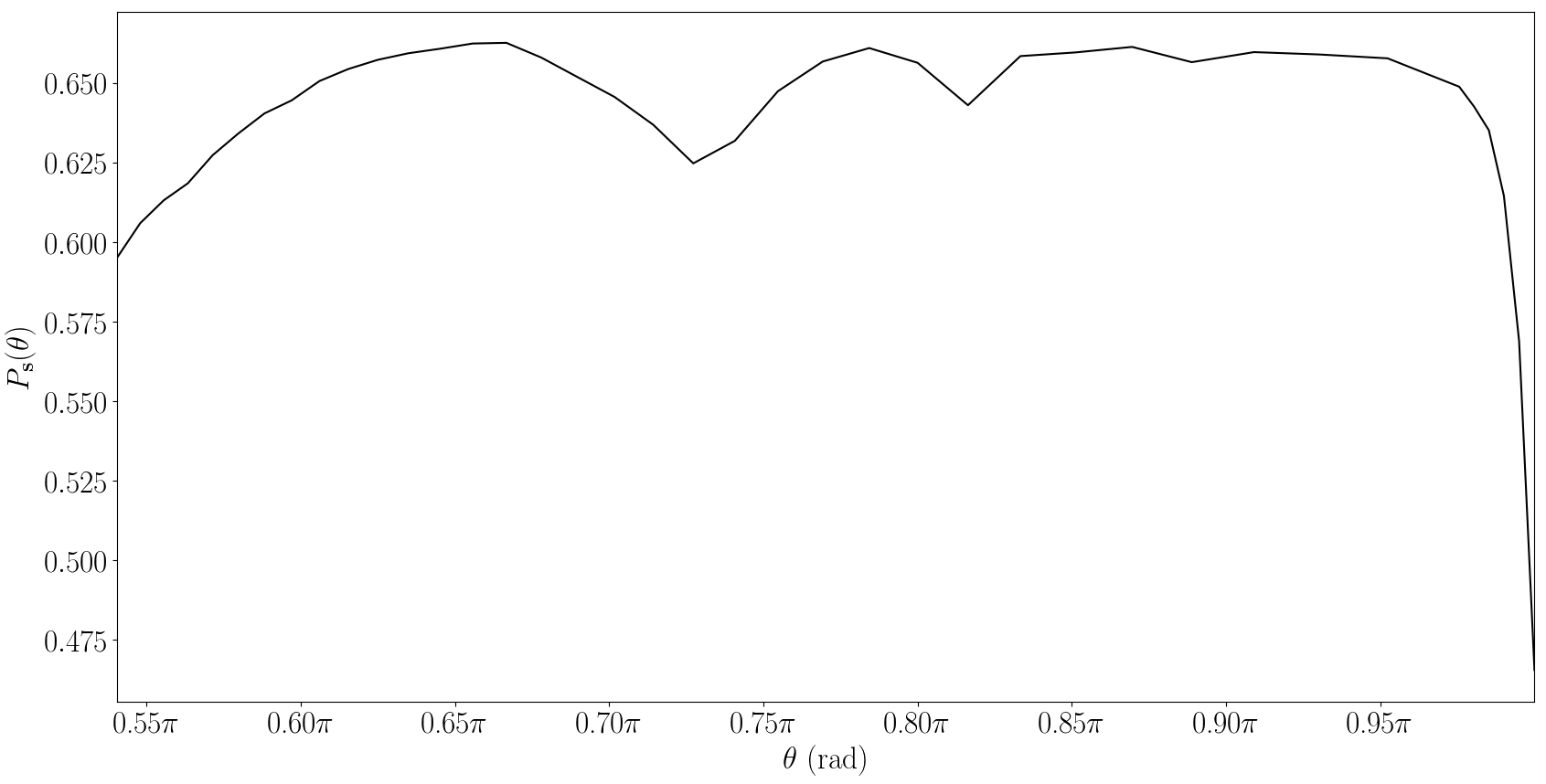}
    \caption{Success probabilities in cogs and stripes regime $0.55\pi\leq\theta\leq \pi $}
    \label{fig:successstripe}
\end{figure}
\begin{figure}
        \begin{subfigure}[b]{0.333\textwidth}
                \includegraphics[width=\linewidth]{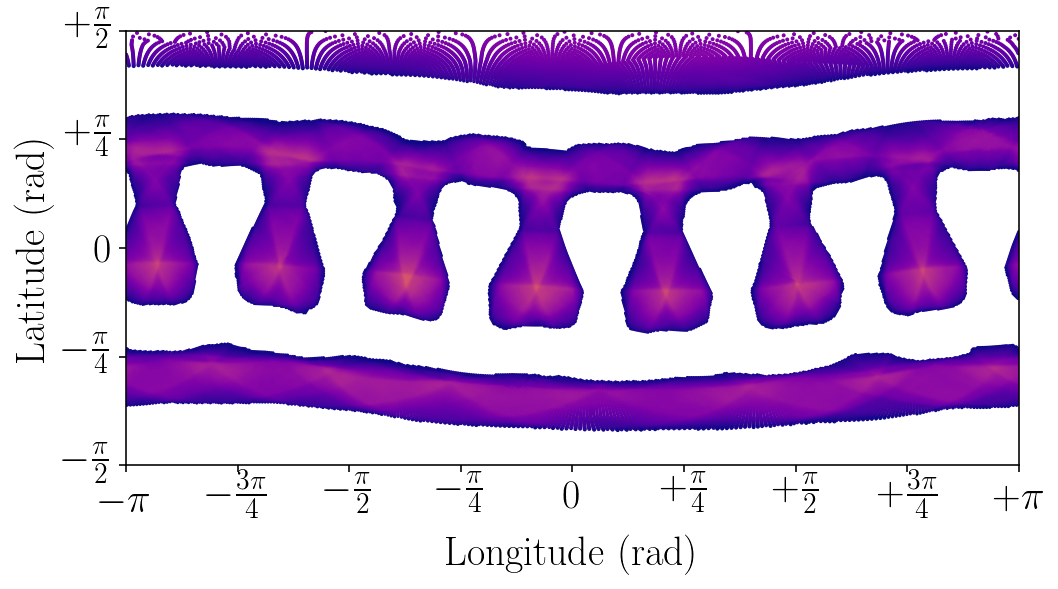}
                \subcaption{$\theta=\frac{2\pi}{3.65}\approx 0.55\pi$}
        \end{subfigure}%
        \begin{subfigure}[b]{0.333\textwidth}
                \includegraphics[width=\linewidth]{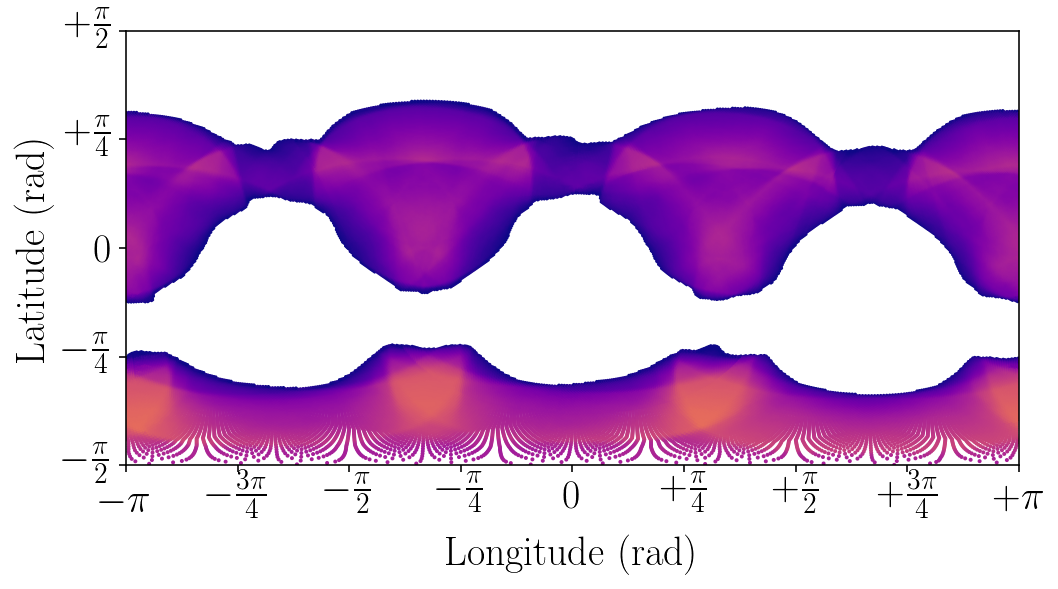} 
                \subcaption{$\theta=\frac{2\pi}{3.5}\approx 0.57\pi$}
        \end{subfigure}%
        \begin{subfigure}[b]{0.333\textwidth}
            \includegraphics[width=\linewidth]{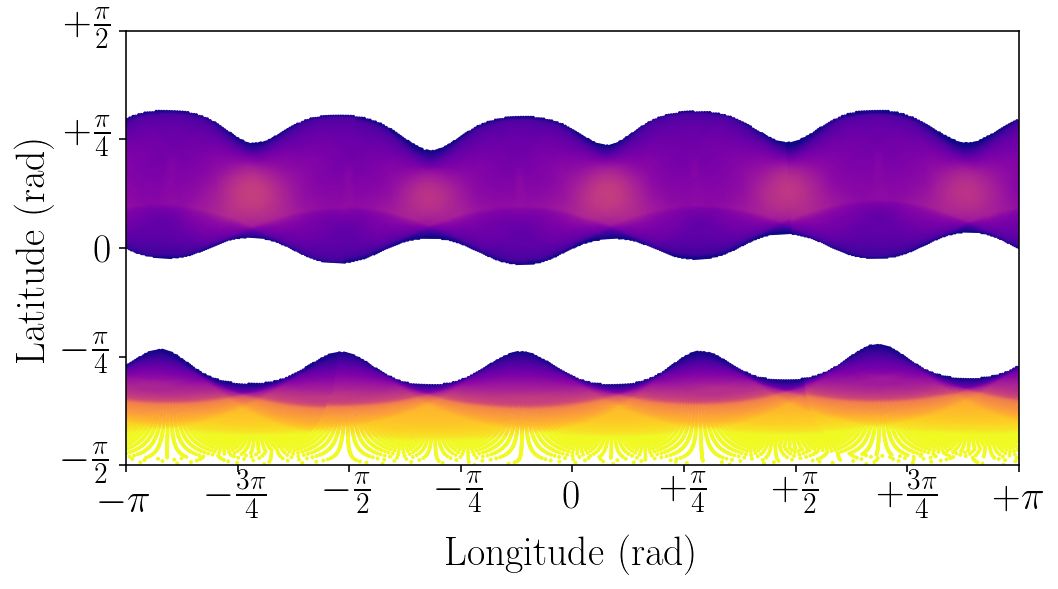}
        \subcaption{$\theta=\frac{2\pi}{3.0}\approx 0.67\pi$}
        \end{subfigure}%
        \hspace{\fill}
        \begin{subfigure}[b]{0.333\textwidth}
                \includegraphics[width=\linewidth]{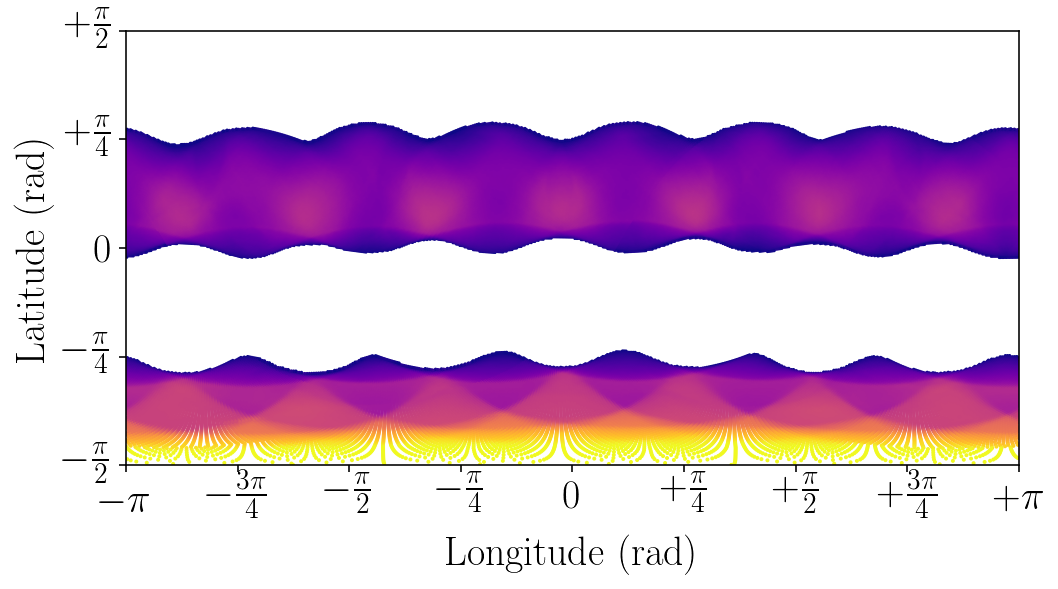}
            \subcaption{$\theta=\frac{2\pi}{2.85}\approx 0.70\pi$}
        \end{subfigure}%
        \begin{subfigure}[b]{0.333\textwidth}
                \includegraphics[width=\linewidth]{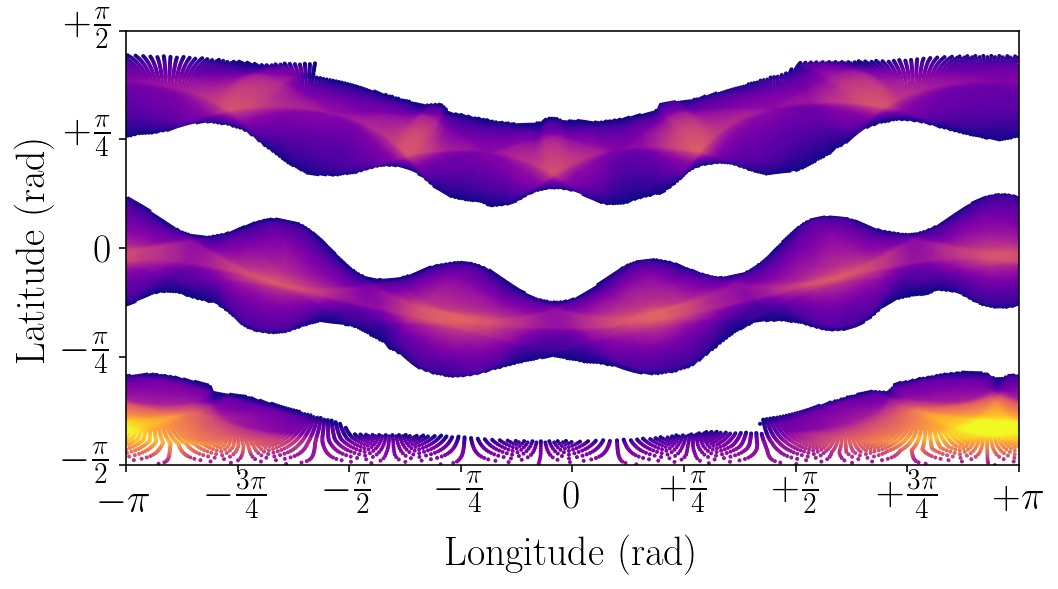}
        \subcaption{$\theta=\frac{2\pi}{2.7}\approx 0.74\pi$}
        \end{subfigure}%
        \begin{subfigure}[b]{0.333\textwidth}
                \includegraphics[width=\linewidth]{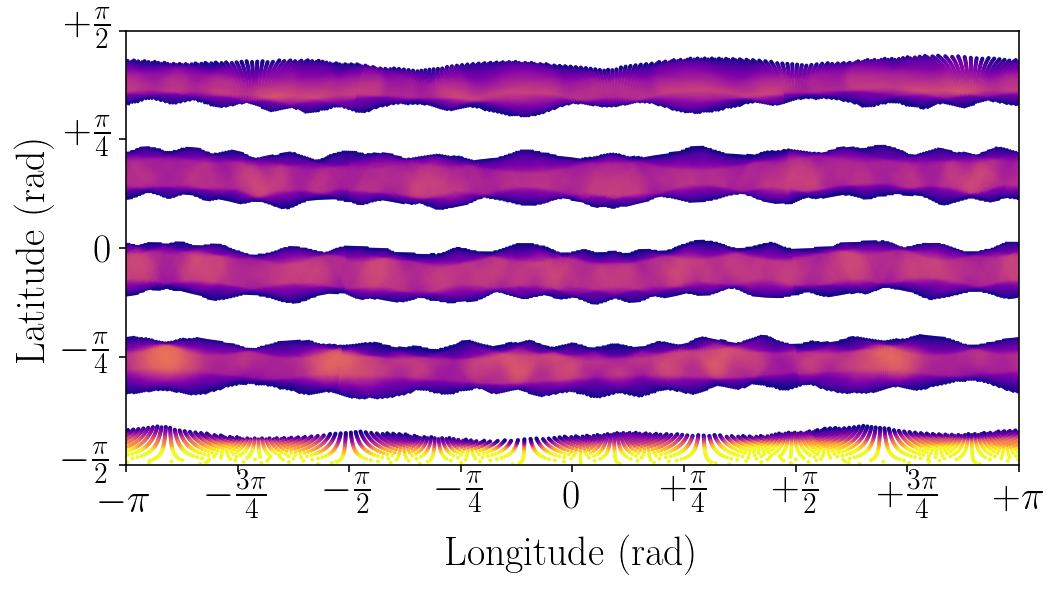}
                \subcaption{$\theta\approx 0.87\pi$}
        \end{subfigure}%
        \hspace{\fill}
        \begin{subfigure}[b]{0.333\textwidth}
                \includegraphics[width=\linewidth]{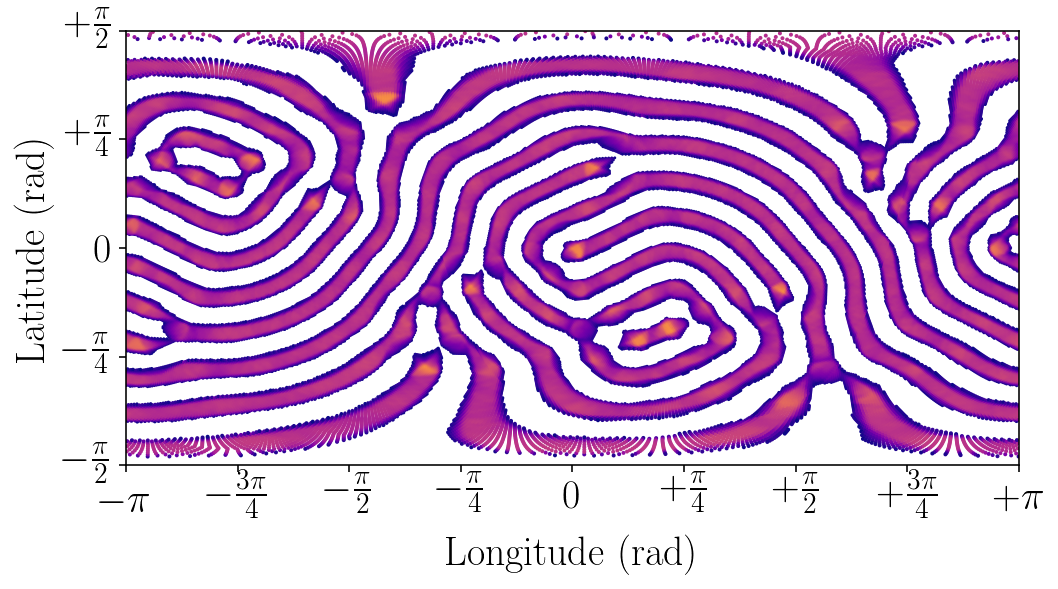}
                \subcaption{$\theta\approx 0.95\pi$}
        \end{subfigure}%
        \begin{subfigure}[b]{0.333\textwidth}
                \includegraphics[width=\linewidth]{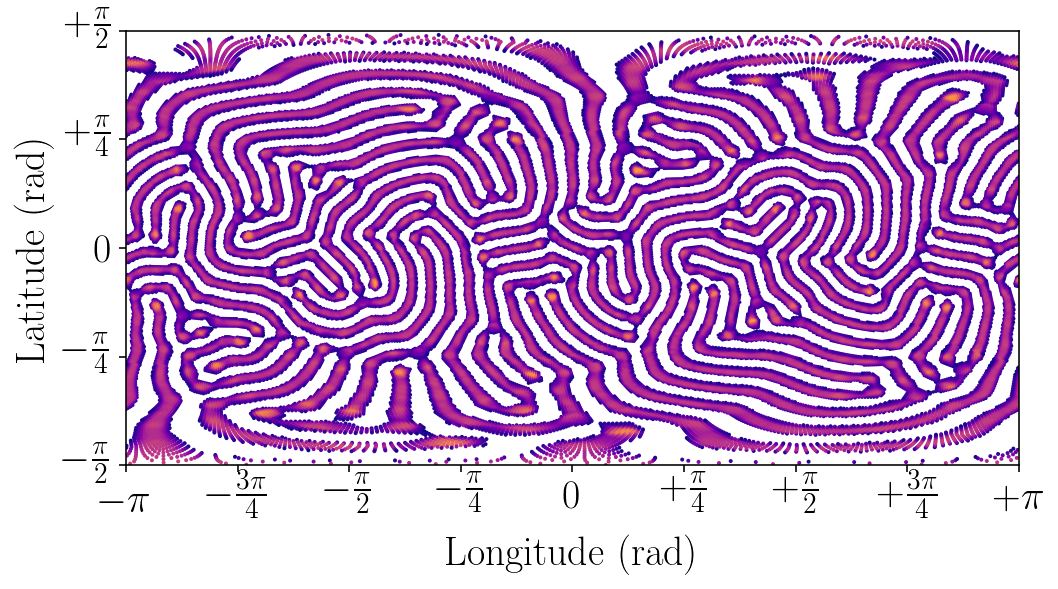}
                \subcaption{$\theta\approx 0.976\pi$}
        \end{subfigure}%
        \begin{subfigure}[b]{0.333\textwidth}
                \includegraphics[width=0.928\linewidth]{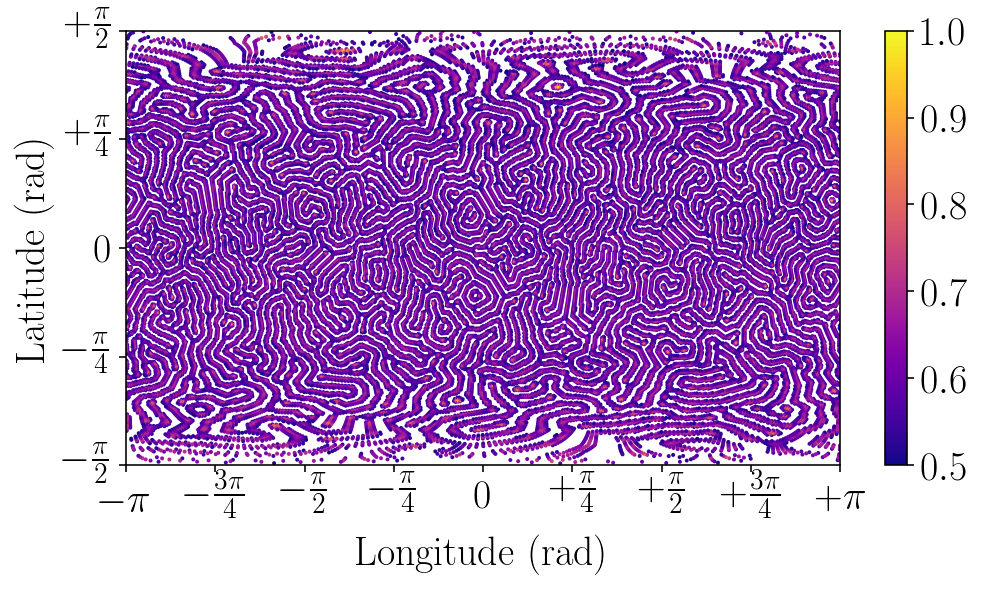}
                \subcaption{$\theta=0.990\pi$}
        \end{subfigure}%
        \caption{Selection of colourings for $\theta$ between $0.55\pi$ and $\pi$. Colour denotes success probability of grid cells and the same colourmap is used for all figures.}
        \label{fig:stripes}
\end{figure}

From $\theta\geq 0.55\pi$, interesting new forms consisting of cogs and stripes are generated. See Figure \ref{fig:successstripe} for the success probabilities and Figure \ref{fig:stripes} for a selection of the colourings. For $0.55\pi\approx\frac{2\pi}{3.55}\leq\theta \leq \frac{2\pi}{3.50}\approx 0.57\pi$ a holed out cogwheel is generated where the number of cogs is 7. The system is found to always converge to this 7-cog cogwheel; a colouring where cogs correlate not with their neighbours, but the next nearest neighbours. For $\theta\geq \frac{2\pi}{3.45}\approx 0.58\pi$ a colourings with 3 cogs is generated. This 3 cog-shape can be regarded as the spherical version of the three-bladed fan from \cite{goulkokent}. 
The difference is that the blades are more circularly shaped and that the three patches around the blades form a pole in the spherical problem. This can also be regarded as a striped configuration as found in \cite{goulkokent}, mapped onto the periodic spherical surface. The circular bulges are also found in the planar stripes, although in the spherical problem these bulges are more prominent. Considering the slow transition from rings with cogs to stripes, we will refer to this regime as the ``cogs and stripes'' regime.  

For $\theta$ between $0.61\pi$ and $0.64\pi$ the extrusions of the 3-cog solution become more irregular, until at $\theta\approx 0.66\pi$ the configuration assumes 5 extrusions. With increasing $\theta$ the extrusions flatten out, until at $0.73\pi$ one is left with a disc at the pole and a band in the other hemisphere. With increasing $\theta$ the number of bands increases; $0.74\pi$ gives a disc at the pole and two bands, $0.82\pi$ gives a pole and three bands. Just like in the cogwheel regime, angles where the colourings change correspond to dips in the success probability, see Figure \ref{fig:successstripe}. From $\theta=0.82$ onwards, the number of stripes increases even more and stripe width decreases. From $0.93\pi$ the bands become irregular and non-circular stripe elements start to appear. Towards $\theta=\pi$ these line elements become thinner and thinner, while the orientation of the lines becomes nonuniform. At $\theta=\pi$ the colouring is uniformly spread over the sphere.

The behaviour in the striped regime is most easily understood by relating the $\theta>\frac{\pi}{2}$ problem to a similar $\theta<\frac{\pi}{2}$ problem. As explained in Sec. \ref{sec:premresults} a correlation between points separated by an angle $\theta\in \left[0,\pi\right]$ is equivalent to a point's antipodal point being anticorrelated with the points at angle $\pi-\theta$. Consequently, for $\theta\geq\frac{\pi}{2}$, the system acts like an Ising model with negative coupling between points separated by angle $\pi-\theta\leq\frac{\pi}{2}$. Put in the context of the grasshopper, the system with $\theta>\frac{\pi}{2}$ converges to the least successful colouring were the angle $\pi-\theta$. This relation explains the generation of stripe colourings and their behaviour towards $\theta=\pi$. With a stripe width that scales with $\pi-\theta$, these lines have a high anti-correlation with the empty spaces in between the stripes. 

Theoretically at angle $\theta=\pi$ the success probability should be zero for any antipodal colouring. However, since the delta function in the correlation is smoothed out over a finite interval, a point's success probability is increased if the point is the same colour as the points separated at an angle close to $\pi$. Using again the relation between maximum colourings for $\theta$ being minimum colourings for $\pi-\theta$, jumping angle $\theta=\pi$ in combination with smoothing of the delta function causes neighbouring states to anticorrelate with each other. This means the system converges to the ground state of an Ising model with negative coupling between neighbours. 

\subsection{Notes on other methods}\label{sec:othermethods}
Noteworthy is that all found solutions can be found using any type of initialisation. In the cogwheel regime, trivially an initialisation of a cogwheel solution with a close number of expected cogs works best. This guarantees that even with a quick cooling scheme simulated annealing does not get stuck in less successful solution like the described double cogs cogwheel or hemisphere. The hemisphere and random initialisation work in the cogwheel regime, but these require a slow cooling procedure. Also interesting is that in general a greedy algorithm performs just as well as simulated annealing. In the transitions of the cogwheel regime the greedy algorithm is prone to getting stuck, especially when a hemisphere colouring is used. Using random or cogwheel initialisation with multiple runs, however, it produces results consistent with simulated annealing. In other regimes the greedy algorithm and simulated annealing have always been found to perform almost equally well. This is further elaborated on in the discussion. 

\section{Discussion and outlook}
The numerical results in this essay have provided qualitative information on the optimal colourings of the grasshopper for the whole range of possible jumping angles $0\leq\theta\leq \pi$. The results are remarkably in line with the planar results found in \cite{goulkokent}. For angles between $0.10\pi$ and $0.41\pi$ cogwheel solutions seem to always be optimal, although in contrast to the planar problem the number of cogs is restricted to odd integers by the antipodality condition. For $\theta$ between $0.41\pi$ and $0.55\pi$ critical colourings are found that can be associated to Ising model behaviour just above the critical point. For $\theta>0.56\pi$ shapes with rings and cogs appear, and these turn into stripes as $\theta$ increases.

\subsection{Resolution}
The smallest angle cogwheel-resembling solution was found for $\theta=\frac{2\pi}{37}\approx 0.05\pi$. For smaller $\theta$ hemisphere solutions are found with irregular boundaries. This can be partly explained by the difference in success probability between a hemisphere solution and a cogwheel solution becoming progressively small with decreasing $\theta$. The energy landscape is therefore very flat, which results in many small steps being needed to escape a local minimum, e.g. the hemisphere solution. Another factor is the resolution. This research used a set-up with $163842$ points, which corresponds to a $h$ of around $0.01$. This means the correlation function is smeared out over a range of $0.04$. For small angles this is not much smaller than the width of a cog, e.g $\frac{\pi}{37}\approx 0.085$ for the 37 cog solution. This resolution flattens out the energy landscape even more, making it harder for the SA algorithm to find the minimum. The SA algorithm is evaluated in Section \ref{sec:SA}.

Achieving a higher resolution is possible by using a finer grid. A compelling idea for this would be using a nonuniform grid with higher resolution around the equator. Using a non-uniform grid would raise challenging questions on how the correlation between points in differently spaced regions is best discretised.

\subsection{Exploration optimisation function}
As the greedy algorithm only allows monotonically increasing success probability, it is interesting that the greedy algorithm performs comparably to the SA approach. This seems to indicate that the underlying optimisation function with the spherical geometry might have an interesting structure.\footnote{Another explanation is that the SA process is malfunctioning and consequently not performing better than the greedy approach. This idea is discussed in Sec. \ref{sec:SA}} Understanding the optimisation function better could provide tools that solve the problem more efficiently, or lead to methods that guarantee a global optimum. For example, an interesting way forward might be finding an (approximate) expression for the shape of the cogs by assuming periodicity around the equator.

\subsection{Improvements SA process} \label{sec:SA}
The simulated annealing process of this essay was not perfect. For small angles $\theta$, it converged to hemisphere solutions and in more difficult settings, as is the case in between cogwheel colourings, it required a lot of steps. Multiple improvements are possible. Firstly, it is worth considering which definition is used for nearest neighbour in the simulated annealing or greedy algorithm. In this essay, a neighbouring state is defined as any state reachable by flipping the colouring of a single antipodal pair of points. This definition is restricting, as it requires a lot of small steps to leave a local minimum. Other definitions might be utilised. An example would be to use \textit{a priori} knowledge of the solution, by also considering as neighbours the states reachable by specific simultaneous flips of a combination of correlated points.\footnote{An example of such a move: for jumping angle close to $\frac{2\pi}{n_o}$, flipping $n_o$ equidistant points on a circle around the z-axis. In this way periodic solutions are easier to reach.}

Another way to make the SA process more efficient is by using continuous time Monte Carlo updates \cite{newman1999monte} when $T$ gets sufficiently small. When $T$ is small, there are only a very small number of points that have a significant probability of being accepted by the Metropolis-Hastings acceptance mechanism (see Sec. \ref{sec:SAmethod}). Continuous time Monte Carlo picks a pair from the pairs that decrease the success probability less than a set threshold, and accepts the flip deterministically. This requires fewer steps in the last part of the cooling process.

Lastly it is possible to use simulated annealing together with parallel tempering \cite{swendsen1986replica}, as also used in \cite{goulkokent}. The combination of these methods reduces the probability the algorithm gets stuck in local optima.

\subsection{Extensions to the Grasshopper Problem}
There are countless ways of generalising the grasshopper problem each of which is interesting to explore in its own right. A first possibility is to drop the antipodality constraint. This research has performed elementary research into the problem without the antipodality condition in the cogwheel regime, this generated colourings consistent with the antipodal findings.\footnote{A trivial difference is, however, that without the antipodality condition cogwheel colourings are not constrained to odd numbers of cogs.} One could further investigate this and make comparisons to the antipodal problem.

The problem can be extended to different spaces and different dimensions, for example platonic solids, higher-dimensional spherical surfaces $\mathbb S^n$ or torus $T^n$. Another option is to extend the problem to a different metric. One could for example explore the cube with an $l^1$ norm. Note that in the discretised case every problem is reduced to a graph with nodes and edges as the cells and correlations, for which one is requested to select a subgraph consisting of exactly half the nodes such that the resulting subgraph has maximum weight (given by the correlations between all points of the subgraph). This means methods for solving the problem could be very similar, only requiring a different grid discretisation and/or different definition for the distance as used for the correlation function. After the correlation matrix between cells is calculated, an identical simulated annealing or greedy algorithm can be used for finding the maximum weighted subgraph.

Another generalisation is changing the way the grasshopper jumps. Forces can be included, such that symmetry in correlation between points is broken. One could also generalise the 1 jump to $N$ jumps, with the objective being either that the grasshopper lands on the grass in the $N$th jump, or also that it cannot jump off the lawn in the first $N-1$ jumps. Additionally, the jumping distance can be made variable. For example, a probability density can be used for the grasshopper's jumping angle. 

Finally, it is worth relating the problem back to the original quantum context. In this essay bounds are sought for the classical correlation function, in the case Alice and Bob use an opposite colouring: the correlation between their measurements for angle $\theta=0$ is -1. A more general problem would be finding two independent colourings; one for Alice and one for Bob. Returning to the analogy of the grasshopper, this could be explained as Alice and Bob needing to seed exactly half of the sphere respectively with grass and flowers, allowing flowers and grass to exist in the same place. Given that the grasshopper lands on any part of the grass (Alice's colouring), the objective is to maximise the probability that after jumping an angle $\theta$, it will land on a flowered patch (Bob's colouring). This essay has explored the specific case that a patch is either covered in both grass and flowers, or in neither. Without this assumption more general Bell inequalities can be explored.

\section{Conclusion}
\begin{figure}
    \centering
    \includegraphics[width=\textwidth]{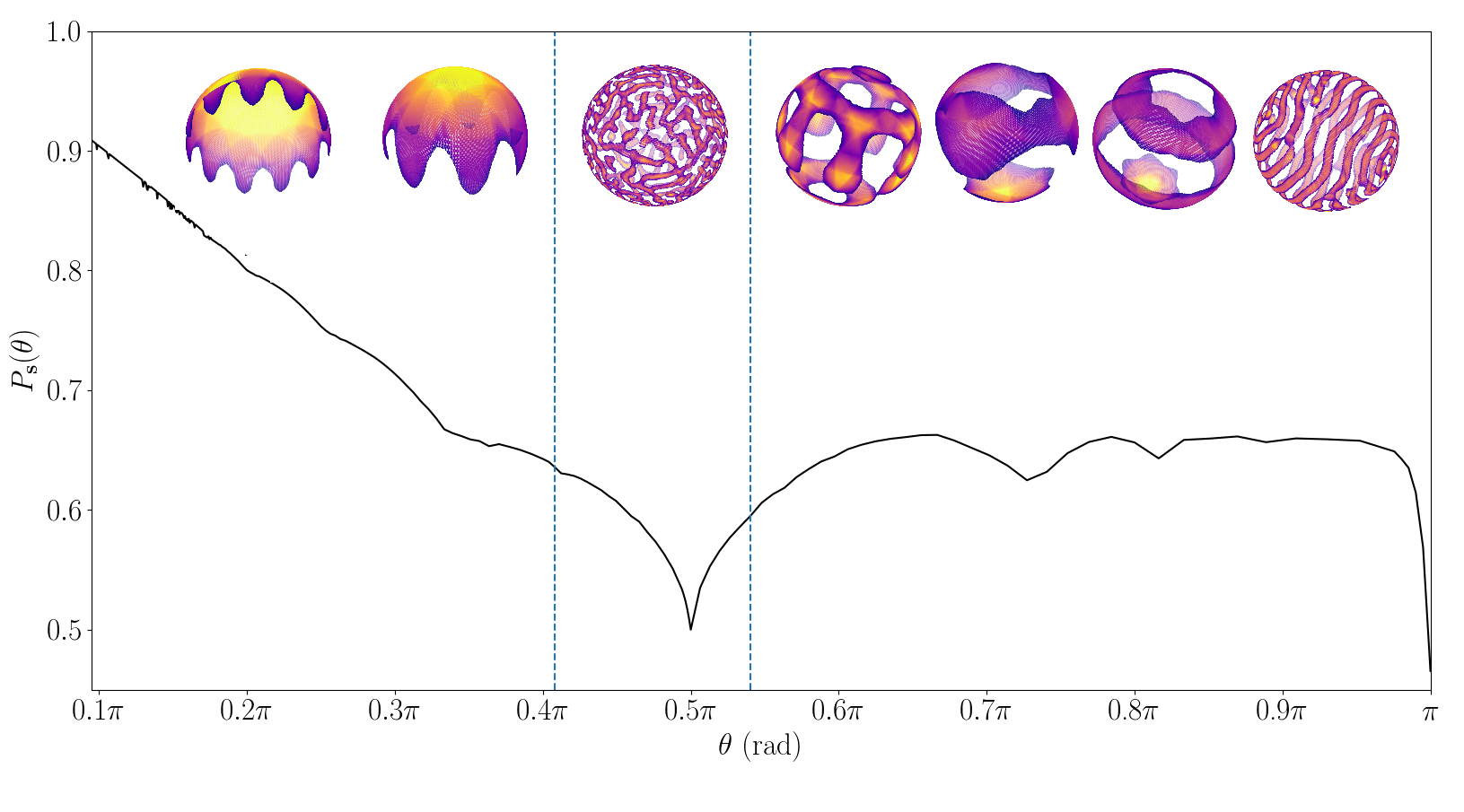}
    \caption{Summary success probability and colouring behaviour. Showing cogwheel regime ($0.10\pi\leq\theta\leq0.41\pi$), critical regime ($0.41\pi\leq\theta\leq0.55\pi$) and cogs and stripes regime ($0.55\pi\leq\theta\leq\pi$).}
    \label{fig:summaryl}
\end{figure}

This research has studied the Grasshopper Problem on the surface of the unit sphere. The problem was discretised, after which a simulated annealing process was used to search for the optimal solution. Results are consistent with the planar results of \cite{goulkokent}. Figure \ref{fig:summaryl} summarises the success probability as a function of $\theta$ and corresponding colouring behaviour. Just like in the planar problem, for small $\theta$ ($0.10\leq\theta\leq 0.41$) cogwheel solutions seem to be optimal. Along the same lines as the planar problem, the regime of cogwheel solutions is followed by a critical regime ($0.41\leq\theta\leq0.55$), where unconnected coloured domains appear that decrease in size as $\theta$ converges to $\frac{\pi}{2}$. The same behaviour in reverse is observed when $\theta$ increases further towards $0.55\pi$. From $\theta=0.55\pi$ the generated colourings resemble holed out cogwheels. For $\theta>0.57$ stripes appear. At first, these stripes display extrusions spaced at angle close to $\theta$, just like in the cogwheel solution. With increasing $\theta$, these extrusions flatten out and the number of stripes increases. Stripe width can be seen to scale with $\pi-\theta$. For $\theta$ close to $\pi$ the stripes become thin and nonuniform in orientation.

For angles smaller than $\frac{2\pi}{25}$ mostly hemisphere solutions with irregular boundaries have been found. Occasionally for small $\theta$ boundaries do display cogs, but these instances are rare and colourings irregular. This is speculated to be the result of a suboptimal simulated annealing implementation, combined with a finite resolution. Improving the simulated annealing process might resolve this problem. Promising possibilities for this are using \textit{a priori} knowledge while considering neighbouring states, using parallel tempering \cite{swendsen1986replica} or using continuous Monte Carlo updates \cite{newman1999monte} for small temperatures. For the studying of set-ups with small angles $\theta$ further research should consider increasing the resolution. An interesting possibility is using a non-uniform grid, although the effects of this on the correlation function have to be taken into account. 

Analysis of the optimisation function in the context of the $\mathbb{S}^2$ geometry is also recommended. In this research a simulated annealing method and greedy algorithm have yielded consistent results for different settings. Better understanding of the optimisation function could explain this behaviour, and offer new tools to tackle the grasshopper problem.

Multiple extensions to the problem have been considered. One could change the problem's space, drop the antipodality constraint or redefine the way the grasshopper jumps. In the context of quantum information further research could study the problem of finding two independent colourings: one for each observer in an EPR-Bohm experiment. This way the Grasshopper Problem could shed light on more general Bell inequalities, inter alia helping to unlock the potential of quantum cryptography.

\clearpage
\section*{Acknowledgements}
This essay was originally written as part of Part III of the Mathemathical Tripos at the University of Cambridge. I would like to warmly thank Adrian Kent, who set the essay and introduced me to this fascinating problem. His suggestions at the start of the project and feedback afterwards were very helpful, and crucial to the project's success.

\bibliographystyle{unsrtnat}
\bibliography{thesis}

\clearpage
\appendix

\section{Non-binary colouring}
\label{sec:appendix}
The Grasshopper problem can be stated using either a binary colouring or a continuous colouring. The latter means that the 1 unit spin per antipodal pair of points can be divided over the two points continuously, so $s_i\in\left[0,1\right]$ instead of $s_i\in\{0,1\}$. We will prove that for the discrete problem the highest possible success probability is achieved by a binary colouring.
\\ \\
\textbf{Lemma} Given an antipodal colouring $\mu$ with non-binary coloured cells $\{\mathbf{v}_i\}$. If antipodal points are not correlated,\footnote{Using the correlation function of this essay this means $\theta\leq\pi-2h$ with $h$ defined as in Eq. \ref{eq:h}} there is a binary colouring $\nu$ with a success probability at least as high as $\mu$.
\\
\\
\textbf{Proof} Given antipodal colouring $\mu$ with non-binary coloured cells $\{\mathbf{v}_i\}$. Let $\mathbf{v}_U$ and $\mathbf{v}_L$ denote an antipodal pair of non-binary coloured cells with zero correlation. Consider the jumping probability of these two cells, defined as the probability a grasshopper will land on black after jumping an angle $\theta$ from the particular cell. We can compare the jumping probability of these two points and shift colour from the cell with the lower probability to the cell with the higher probability. In the case both cells have equal success probability $0.5$, the direction of colour shift is arbitrary. This process results in one cell coloured and one cell uncoloured. By keeping the colouring of all the other cells constant, this operation causes a linear increase in the total success probability, see Eq. \ref{eq:success}. 

This operation can be repeated for every pair of non-binary coloured cells until a binary colouring $\nu$ is achieved, every time resulting in a non-negative change in total success probability. Conclusion: for every non-binary colouring $\mu$ a nearby binary colouring $\nu$ can be found with higher (or identical) success probability.


\end{document}